\documentclass[a4paper,11pt]{article}
\pdfoutput=1
\usepackage{jcappub}

\usepackage{fontawesome}
\usepackage{blkarray}
\usepackage{graphicx}
\usepackage{amsfonts}
\usepackage{soul}
\usepackage{amssymb}
\usepackage{amsmath}
\usepackage{pifont}
\usepackage{slashed} 
\usepackage{enumerate}
\usepackage{color}
\usepackage{adjustbox}
\usepackage{comment}
\usepackage{bbold}
\usepackage{tikz}
\usepackage{amsmath}
\usepackage{tikz}
\usetikzlibrary{tikzmark}
\usepackage{empheq}
\usepackage{adjustbox}
\usepackage{multirow}
\usepackage[utf8]{inputenc}
\usepackage{adjustbox}

\usepackage{tikz-feynman}
\usepackage{tcolorbox}

\usepackage{enumitem}

\definecolor{oucrimsonred}{rgb}{0.6, 0.0, 0.0}
\definecolor{persianblue}{rgb}{0.11, 0.22, 0.73}
\definecolor{forestgreen}{rgb}{0.13,0.35,0.13}
\definecolor{lightgray}{rgb}{0.83, 0.83, 0.83}
\definecolor{cornellred}{rgb}{0.0, 0.0, 0.5}
\definecolor{amethyst}{rgb}{0.6, 0.4, 0.8}
\definecolor{yellow}{rgb}{1.0, 1.0, 0.0}
\definecolor{firebrick}{rgb}{0.7, 0.13, 0.13}
\definecolor{tangerineyellow}{rgb}{1.0, 0.8, 0.0}
\definecolor{deepfuchsia}{rgb}{0.76, 0.33, 0.76}
\definecolor{amber}{rgb}{1.0, 0.75, 0.0}
\definecolor{VioletRed4}{rgb}{0.55, 0.13, .32}
\definecolor{indiagreen}{rgb}{0.07, 0.53, 0.03}
\definecolor{VioletRed4}{rgb}{0.55, 0.13, .32}
\newcommand{\nn}{\nonumber}

\definecolor{oucrimsonred}{rgb}{0.6, 0.0, 0.0}
\newcommand\vertarrowbox[3][6ex]{%
  \begin{array}[t]{@{}c@{}} #2 \\
  \left\uparrow\vcenter{\hrule height #1}\right.\kern-\nulldelimiterspace\\
  \makebox[0pt]{\scriptsize#3}
  \end{array}%
}

\definecolor{mtcolor}{rgb}{.8,.3,.1}

\definecolor{violachiaro}{rgb}{1,0.6,1}
 
\definecolor{gbcolor}{rgb}{.43,.22,.12}
 
\definecolor{gbcolor2}{rgb}{.9,.2,.6}
\definecolor{gbcolor3}{rgb}{.3,.2,.6}

\definecolor{verdechiaro}{rgb}{0.6,1,0.6}
\definecolor{giallochiaro}{rgb}{1,1,0.6}
\definecolor{bluscuro}{rgb}{0.15, 0.2, 0.9}
\definecolor{verdes}{rgb}{0.1, 0.5, 0.1}%
\definecolor{tangerineyellow}{rgb}{1.0, 0.8, 0.0}
\definecolor{americanrose}{rgb}{1.0, 0.01, 0.24}
\definecolor{cobalt}{rgb}{0.0, 0.28, 0.67}
\definecolor{brandeisblue}{rgb}{0.0, 0.44, 1.0}
\definecolor{mycolor}{rgb}{0.0, 0.0, 0.5}
\definecolor{oxfordblue}{rgb}{0.0, 0.13, 0.28}
\definecolor{azure}{rgb}{0.0, 0.5, 1.0}
 
\definecolor{turquoiseblue}{rgb}{0.0, 1.0, 0.94}
\newtcolorbox{mynewbox}[1]{colback=white!5!white,colframe=azure!75!black,fonttitle=\bfseries,title=#1}
\newtcolorbox{mybox}{colback=mycolor!5!white,colframe=azure!75!black}
\newtcolorbox{mynamedbox}[1]{colback=mycolor!5!white,colframe=azure!75!black,title=#1}
\definecolor{venetianred}{rgb}{0.78, 0.03, 0.08}
\newtcolorbox{mynamedbox2}[1]{colback=venetianred!5!white,colframe=venetianred!80!black,title=#1}

\definecolor{myforestgreen}{rgb}{0.13, 0.55, 0.13}
\definecolor{rossocorsa}{rgb}{0.83, 0.0, 0.0}
\definecolor{rossocorsa2}{rgb}{0.75, 0.0, 0.0}

\makeatletter
\def\l@subsubsection#1#2{}
\makeatother

\definecolor{oucrimsonred}{rgb}{0.6, 0.0, 0.0}
\definecolor{persianblue}{rgb}{0.11, 0.22, 0.73}
\definecolor{forestgreen}{rgb}{0.13,0.35,0.13}

\newcommand{\lupm}{LUPM, CNRS, Université Montpellier Place Eugene Bataillon, F-34095 Montpellier, France}
\newcommand{\uniroma}{Dipartimento di Fisica, ``Sapienza'' Universit\`a di Roma, Piazzale Aldo Moro 5, 00185, Roma, Italy}
\newcommand{\infn}{INFN sezione di Roma, Piazzale Aldo Moro 5, 00185, Roma, Italy}
\newcommand{\NICPB}{Laboratory of High Energy and Computational Physics, National Institute of Chemical Physics and Biophysics, R\"{a}vala pst. 10, 10143, Tallinn, Estonia}

\title{
Primordial black holes in the curvaton model: possible connections to pulsar timing arrays and dark matter 
}

\author[a,b]{Giacomo Ferrante,}
\emailAdd{giacomo.ferrante@umontpellier.fr}
\author[b,c]{Gabriele Franciolini,}
\emailAdd{gabriele.franciolini@uniroma1.it}
\author[b,c,d]{Antonio Junior Iovino}
\emailAdd{antoniojunior.iovino@uniroma1.it}
\author[b,c]{and Alfredo Urbano}
\emailAdd{alfredo.urbano@uniroma1.it}

\affiliation[a]{\lupm}
\affiliation[b]{\uniroma}
\affiliation[c]{\infn}
\affiliation[d]{\NICPB}

\abstract{
We revise primordial black holes (PBHs) production in the axion-curvaton model, 
in light of recent developments in the computation of  their abundance accounting for non-gaussianities (NGs) in the curvature perturbation up to all orders. 
We find that NGs intrinsically generated in such scenarios have a relevant impact on the phenomenology associated to PBHs and, in particular, on the relation between the abundance and the signal of second-order gravitational waves. 
We show that this model could explain both the totality of dark matter in the asteroid mass range and the tentative signal reported by the NANOGrav and IPTA collaborations in the nano-Hz frequency range.
En route, we provide a new, explicit computation of the power spectrum of curvature perturbations going beyond the sudden-decay approximation.
}

\begin{document}
\maketitle
\flushbottom

\section{Introduction}\label{sec:Summary}

The idea of black holes produced in the early Universe was firstly proposed by Novikov and Zel'Dovic\,\cite{Zeldovich:1967lct}, followed shortly by the works by Hawking and Carr\,\cite{Hawking:1971ei, Carr:1974nx}, and has recently attracted much attention as they may contribute to the gravitational wave (GW) detections reported by the LIGO/Virgo/KAGRA collaboration~(e.g.\,\cite{Bird:2016dcv,Sasaki:2016jop,Clesse:2016vqa, Ali-Haimoud:2017rtz, Raidal:2018bbj,Franciolini:2021tla,Liu:2021jnw,Franciolini:2022tfm,Escriva:2022bwe}). 
Furthermore, PBHs are also currently in the spotlight as they could explain the totality of dark matter 
(or at least a sizeable fraction of it) while being unavoidably related to a signal of second-order GWs, potentially visible at future GW experiments.
It is of particular interest that the induced GW signatures of the formation of stellar mass PBHs fall within the reach of current Pulsar Timing Array (PTA) experiments, in the nano-Hz frequency range, 
where the NANOGrav Collaboration recently reported evidence for a stochastic common process~\cite{NANOGrav:2020bcs} (and also independently supported by other PTA data\,\cite{Goncharov:2021oub,Chen:2021rqp,Antoniadis:2022pcn}).
Also, at larger frequencies, LISA\,\cite{LISACosmologyWorkingGroup:2022jok} would be able to observe imprints of asteroidal mass PBH formation\,(e.g. \cite{Saito:2008jc,Garcia-Bellido:2017aan,Cai:2018dig,Bartolo:2018evs}). 

Different PBH formation mechanisms were devised in the literature. In this work, we will focus on what is denoted as the standard scenario in which PBH formation occurs by gravitational collapse of sufficiently large-amplitude perturbations in the density field (see e.g.~\cite{Sasaki:2018dmp} for a review). 
In recent years, numerous studies have shown how a spectator field, known as curvaton, can lead to the creation of large-amplitude perturbations during inflation\,\cite{Enqvist:2001zp,Lyth:2001nq,Sloth:2002xn,Lyth:2002my,Dimopoulos:2003ii}. Under certain conditions, these perturbations can be responsible for the generation of PBHs\,\cite{Kohri:2012yw,Kawasaki:2012wr,Kawasaki:2013xsa,Bugaev:2013vba,Ando:2017veq,Ando:2018nge,Chen:2019zza,Inomata:2020xad,Liu:2020zzv,
Pi:2021dft,
Liu:2021rgq,
Kawasaki:2021ycf}.

We study PBH production in the frame of a specific axion-like curvaton model\,\cite{Kawasaki:2012wr,Ando:2017veq,Inomata:2020xad}, where a complex scalar field $\Phi=\varphi e^{i \vartheta}$ is introduced and its angular component $\vartheta$ is dubbed \textit{curvaton}.
The rolling of the radial component $\varphi$ down the potential during inflation enhances by many orders of magnitude the angular perturbations $\delta \theta$ at small scales, while, 
at scales associated with Cosmic Microwave Background (CMB) observations, $\delta \theta$ is suppressed and the main contribution to density fluctuations comes from the inflaton field, which lives in an uncoupled sector.
Exploiting these features, we are able to obtain a broad power spectrum of curvature perturbations, i.e. a power spectrum which spans over many orders of magnitude of scales $k$, in a concrete model of the early universe. 
As it will be shown in section\,\ref{sec:Pheno}, this is crucial in order to make connection between observables related to PBH dark matter and those associated with scalar-induced GWs. 

As already discussed at length in the literature, NGs of the curvature perturbation $\zeta$ 
play a major role in the computation of the abundance (see e.g.\,\cite{Young:2013oia,Franciolini:2018vbk}) for multiple reasons. 
First of all, the criterion for PBH collapse is expressed in terms of a critical density contrast threshold.
This requires computing the abundance by deriving the statistical distribution of overdensity perturbations $\delta$, which in turn are non-linearly related to $\zeta$ at super-Hubble scales (see e.g.\,\cite{Harada:2015yda}). 
This non-linear relation unavoidably introduces NGs in the distribution of the curvature perturbation\,\cite{DeLuca:2019qsy,Young:2019yug}, which we will refer to in the following as \textit{NGs from non-linearities}. Furthermore, curvaton dynamics leads to NG corrections in the probability density function (PDF) of $\zeta$ itself\,\cite{Bartolo:2003jx,
Bartolo:2005fp,
Sasaki:2006kq,
Enqvist:2008gk,
Kohri:2009ac,
Chingangbam:2009xi,
Huang:2010cy,
Kawasaki:2011pd,Fonseca:2011aa,Kawasaki:2012gg,Enomoto:2012uy,
Mukaida:2014wma,
Liu:2020zlr,
Ghoshal:2023lly}. Such corrections, denoted as \textit{primordial NGs}, are usually taken into account by means of the perturbative expansion
\begin{equation}
\label{eq:FirstExpansion}
    \zeta=\zeta_{\rm G}+\frac{3}{5}f_{\rm NL}\zeta_{\rm G}^2+\frac{9}{25}g_{\rm NL}\zeta_
{\rm G}^3+\dots\,,
\end{equation}
where $\zeta_{\rm G}$ follows a Gaussian distribution. The above expansion is often truncated at the quadratic order, assuming that $f_{\rm NL}$ captures the leading non-gaussian correction. 
However, the coefficients $f_{\rm NL}, g_{\rm NL},\dots$ are generally not independent and, in the case of the curvaton model, can be resummed into an expression of the form $\zeta=\log\left[X(\zeta_{\rm G})\right]$\,\cite{PhysRevD.74.103003}, where $X$ will be introduced in section \ref{sub:X}. 
Most importantly, in the case of a broad power spectrum, the computation of the abundance requires evaluating the power-series expansion in eq.\,(\ref{eq:FirstExpansion}) beyond its radius of convergence, and thus leads to a mathematically flawed result, as recently pointed out in ref.\,\cite{PhysRevD.107.043520}.
In this work, we apply the technique developed in ref.\,\cite{PhysRevD.107.043520} to the aforementioned curvaton model (see also ref.~\cite{Gow:2022jfb}) and compute abundance of PBHs by considering both intrinsic and NGs induced by non-linearities in a non-perturbative way. 
As we will see, this has important consequences for the predicted signal of scalar-induced GWs, and in particular, allows to fit within a simple curvaton scenario both PBH dark matter and a stochastic GW background (SGWB) compatible with NANOGrav and IPTA recent data.

The paper is organized as follows. In section\,\ref{sec:TypeI} we study in details the axion-like curvaton model, showing how it is able to produce a broad spectrum of curvature perturbations, going beyond the sudden decay approximation.  
The major results of our work are presented in section\,\ref{sec:Pheno}, where we compute the observables related to the model taking into account all the sources of NGs. These are the PBH mass distribution and the associated SGWB. Conclusions and discussion of results are presented in section\,\ref{sec:Finale}.

\section{Axion-like curvaton model}
\label{sec:TypeI}

\subsection{The inflationary dynamics}
For simplicity, we work in the limit in which the Hubble rate during inflation is constant (pure de Sitter background), 
and we indicate its value with $H_{\rm inf}$. 
We consider the axion-like curvaton model explored in refs.\,\cite{Kawasaki:2012wr,
Ando:2017veq,
Inomata:2020xad}.  
In these models the radial field is subject to the quadratic potential
\begin{align}\label{eq:EffectiveV}
V(\varphi) = \frac{c}{2}H_{\rm inf}^2(\varphi - f)^2\,,
\end{align}
which has a minimum at $\varphi = f$. 
Refs.\,\cite{Dine:1995kz,
Kawasaki:2012wr,
Ando:2017veq,
Inomata:2020xad} justify the potential in eq.\,(\ref{eq:EffectiveV}) in the context of supergravity models.

If the radial field $\varphi$ during inflation rolls down the quadratic potential in eq.\,(\ref{eq:EffectiveV}), starting from
some large field value of Planckian order,  the angular perturbations get a large enhancement. This is easy to see explicitly. 
The equation of motion for $\varphi$ (written in terms of the number of $e$-folds $N$, defined by $dN = 
H dt$, as time variable)
\begin{align}
\frac{d^2\varphi}{dN^2} + 3\frac{d\varphi}{dN} + \frac{1}{H_{\rm inf}^2}\frac{dV(\varphi)}{d\varphi} =0\,,
\end{align}
admits the analytical solution 
\begin{align}\label{eq:NaiveSolution}
\varphi_H(N) = f_H + c_1 e^{N(-3 - \sqrt{9-4c})/2} + c_2 e^{N(-3 + \sqrt{9-4c})/2}\,,
\end{align}
where we introduce the a-dimensional field $\varphi_H \equiv \varphi/H_{\rm inf}$ 
and we define $f_H \equiv f/H_{\rm inf}$; the constants $c_{1,2}$ are fixed by some initial condition 
$\varphi_H(N_*) = \varphi_*$ and, for simplicity, vanishing initial velocity. 
We consider the case in which $0 < c < 9/4$.
In this case, the field value in eq.\,(\ref{eq:NaiveSolution}) decreases exponentially fast starting from the value $\varphi_*$.

From the above expression we can easily compute the value $\varphi_H(N_k)$, that is 
the value of the field $\varphi_H$ at the time $N_k$ at which the mode with comoving wavenumber $k$ crosses the Hubble horizon
 $a(N_k)H = k$. We define
\begin{align}\label{eq:HorizonExit}
N_k = N_* + \log\left (\frac{k}{k_*} \right)\,,
\end{align}
where $k_*$ is the comoving wavenumber that crosses the horizon at time $N_*$, that is $a(N_*)H = k_*$. 
To fix ideas, we consider $k_* = 0.05$ Mpc$^{-1}$ as a reference scale with $N_* = \mathcal{O}(60)$.
We find
\begin{align}
\varphi_H(N_k) &= \frac{1}{2\bar{c}}\bigg(\frac{k}{k_*}\bigg)^{-(3+\bar{c})/2}\Bigg\{
\varphi_*\left[
-3+\bar{c}+(3+\bar{c})\bigg(\frac{k}{k_*}\bigg)^{\bar{c}}
\right] 
\nonumber 
\\
&- f_H\left[
-3+\bar{c} - 2\bar{c}\bigg(\frac{k}{k_*}\bigg)^{(3+\bar{c})/2} + (3+\bar{c})\bigg(\frac{k}{k_*}\bigg)^{\bar{c}}
\right]
\Bigg\}\,,
\end{align}
with $\bar{c}\equiv \sqrt{9-4c}$. 
This expression is important because it shows how the factor $1/|\varphi_H(N_k)|^2$ enters in the determination of the amplitude of the angular perturbations once they exit the 
horizon. Consequently, one finds the analytical result  
\begin{align}\label{eq:AngularPert}
k^{3/2}|\delta\vartheta_k| = \frac{1}{\sqrt{2}\varphi_H(N_k)}~~~~~~~~~
\Longrightarrow~~~~~~~~~
P_{\delta\theta}(k) = \frac{k^3|\delta\vartheta_k|^2}{2\pi^2} 
= \frac{1}{4\pi^2|\varphi_H(N_k)|^2}\,.
\end{align} 
The angular power spectrum grows as a power-law 
$P_{\delta\theta}(k) \propto (k/k_*)^{n_{\theta}}$ with spectral index given by 
$n_{\theta} = 3 - \sqrt{9-4c}$. 
If the field $\varphi$ rolls from Planckian values down to  $\varphi_H = \mathcal{O}(f_H)$, one gets many orders-of-magnitude of 
power-law enhancement which is eventually crucial for the formation of PBHs or the 
generation of a sizable GW signal. 
More concretely, the power spectrum  of angular fluctuations ranges in between the two limiting values
\begin{align}\label{eq:PowerSpectrumRange}
 \frac{1}{4\pi^2 \varphi_*^2}~[{\rm for\,} k/k_*=1]~
 \leqslant~P_{\delta\theta}(k)~\leqslant~\frac{1}{4\pi^2 f_H^2}~[{\rm for\,} k/k_* \gg 1]
 ~~~\Longrightarrow~~~
 \Delta P_{\delta\theta} \equiv \frac{\varphi_*^2}{f_H^2}\,,
\end{align}
so that the  enhancement is controlled 
by the ratio $\varphi_*/f_H$.

Another relevant information is the time it takes for the power spectrum to fully grow from its initial value 
at $k/k_* = 1$
up to $1/4\pi^2 f_H^2$. 
Let us define this quantity as $\Delta N$. If we approximate the power spectrum with a power-law (since we are only interested 
in the growing part), we find
\begin{align}
P_{\delta\theta}(k) \approx \frac{1}{4\pi^2\varphi_*^2}\left(
\frac{k}{k_*}
\right)^{3 - \sqrt{9-4c}}
\end{align}
which leads to 
\begin{align}\label{eq:TimeScaling}
\Delta N = \frac{1}{3-\sqrt{9-4c}}\log\left(\frac{\varphi_*^2}{f_H^2}\right) = 
\frac{\log(\Delta P_{\delta\theta})}{3-\sqrt{9-4c}}\,.
\end{align}
The result depends on $c$ since it controls the slope of the growing power spectrum.

\begin{figure}[!htb!]
\begin{center}
$$\includegraphics[width=.49\textwidth]{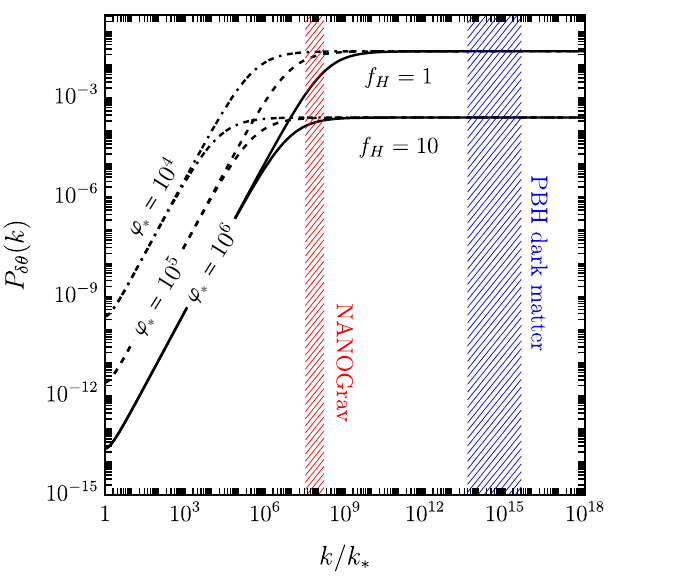}
\quad\includegraphics[width=.49\textwidth]{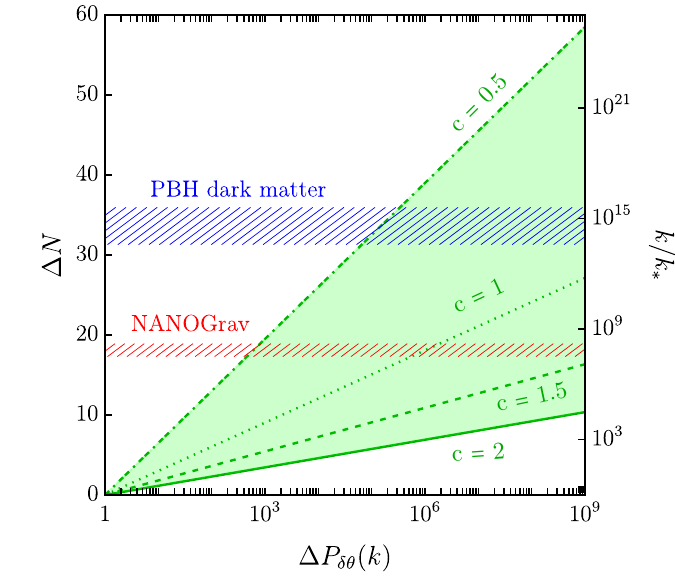}$$
\caption{\em \label{fig:PSboosted} 
Left panel: 
angular power spectrum for different choices of $\varphi_*$ and $f_H$ and fixed 
$c=27/16$ 
(leading to a spectral growth $n_\theta =1.5 $).
Right panel:
different spectral growth as a function of $\Delta N$ depending on the assumed initial conditions for the radial field dynamics, i.e. different $c$.
For illustration, in both panels, we show the frequency range (converted in units of $k/k_*$) that is relevant for NANOGrav and other PTA experiments and the mass range (converted in units of $k/k_*$) in which 
the totality of the present-day dark matter abundance 
can be compatible with the PBH hypothesis (see main text for details).
 }
\end{center}
\end{figure}

We show the power spectrum in eq.\,(\ref{eq:AngularPert}) in the left panel of fig.\,\ref{fig:PSboosted} for 
fixed $c$ but different choices of $\varphi_*$ and $f_H$ as function of $k/k_*$ (see caption for details). In the right panel of the same figure, we plot the function $\Delta N$ in eq.\,(\ref{eq:TimeScaling}) 
as function of $\Delta P_{\delta\theta}$ for different values of $c$.
For illustration we superimpose {\it i)} in red, the range of frequencies $2.5\times 10^{-9} \lesssim f\,\,[{\rm Hz}] \lesssim  1.2\times 10^{-8}$
that is relevant for the observation of a potential SGWB signatures by the NANOGrav experiment\,\cite{NANOGrav:2020bcs} (and other PTAs \cite{Goncharov:2021oub,Chen:2021rqp,Antoniadis:2022pcn}) and {\it ii)} in blue the range of mass in which PBHs may constitute the totality of dark matter observed in the present-day Universe (given existing constraints  \cite{Carr:2020gox}), $10^{18} \lesssim M_{\rm PBH}\,[{\rm g}] \lesssim 10^{21}$.

Starting from ad-hoc initial conditions to get the above result, i.e. $\varphi_*$ of planckian order, is not entirely satisfactory and we address the degree of fine tuning required for setting our initial conditions in appendix\,\ref{app:Ini}.

\begin{figure}[!htb!]
\begin{center}
\includegraphics[width=.85\textwidth]{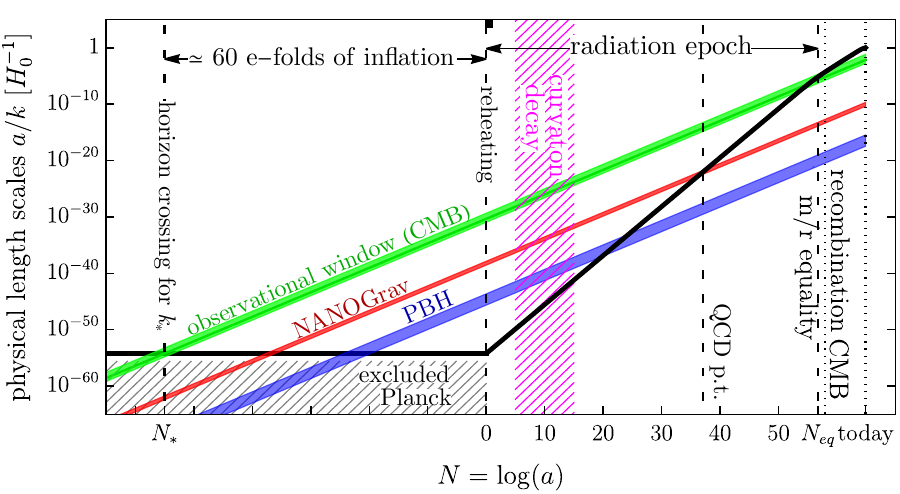}
\caption{\em \label{fig:TaleOfScales}The black line shows the evolution of the physical Hubble horizon $1/H$ as a function of the number of $e$-folds $N$, computed assuming slow-roll inflation and standard $\Lambda$CDM model. The green region shows the evolution of the scales associated with CMB observations. The scales of phenomenological interest for PBHs are those associated with a signal of second-order GWs compatible with NANOGrav and PTA data (red region) and those at which PBHs account for the totality of dark matter (blue region). In both cases it is crucial that such scales re-enter the Hubble horizon after all the curvaton field has decayed into radiation otherwise PBHs production cannot occur by means of gravitational collapse of the radiation fluid, as expected in curvaton models. Interestingly enough, the horizon crossing of scales relevant for a detection by the NANOGrav collaboration occurs around the time of QCD phase transition. This means that the phase transition due to the confinement of the strong sector is of phenomenological relevance and has to be taken into account when computing observables of the model.}
\end{center}
\end{figure}

At this stage, we can already make a number of relevant comments.  
\begin{itemize}
\item[{\it i)}]
The power spectrum of angular perturbations $\delta\vartheta_k$ can be enhanced during inflation -- triggered by the 
classical rolling of its radial counterpart -- by several orders of magnitude. The key equation is eq.\,(\ref{eq:TimeScaling}).  
The a-dimensional ratio $\Delta P_{\delta\theta} \equiv \varphi_*^2/f_H^2$ controls the 
enhancement of the power spectrum and the parameter $c$ controls the $e$-fold time 
$\Delta N$ that is needed to complete such a growth, see fig.\,\ref{fig:PSboosted}. 
This is because the larger $c$ the faster the classical rolling of the radial field (see eq.\,(\ref{eq:NaiveSolution})). 
In particular, we remark that, as follows from the right-hand side of the inequality in 
eq.\,(\ref{eq:PowerSpectrumRange}), the smaller $f_H$ the larger the upper value reachable by   
$P_{\delta\theta}(k)$.
\item[{\it ii)}] By requiring the condition $9-4c>0$, we  consequently fix the maximum spectral growth to $3$ and we can see from the left  panel of fig.\,\ref{fig:PSboosted} that $P_{\delta \theta}$ does not exhibit oscillatory behavior when reaching the plateau.
In principle, violating the condition $9-4c>0$ would cause $\varphi$ to reach the minimum of the quadratic potential and oscillate around it, thereby altering the shape of the angular power spectrum as described by eq.\,(\ref{eq:AngularPert}). 
In our case, this does not occur due to the dominant role of the Hubble friction term, which brings $\varphi$ to $f$ asymptotically and damps any oscillations.
As a result, we will not generate any significant oscillatory features on the PBH abundance nor on the SGWB spectrum
\item[{\it iii)}]
In fig.\,\ref{fig:TaleOfScales} we show the evolution of physical length-scales $\lambda_{\rm phys} = a/k$ 
(in units of the present-day Hubble length $1/H_0$) throughout 
the history of the Universe from the inflationary epoch until today. 
We assume instantaneous reheating and, after inflation, standard $\Lambda$CDM cosmology.
Experimental constraint on the tensor-to-scalar ratio can be translated into an upper bound on the energy scale of inflation; in turn, this implies an upper bound on the Hubble parameter during inflation\,\cite{Planck:2018jri}
\begin{align}
\frac{H_{\rm inf}}{\bar{M}_{\rm Pl}} < 2.5\times 10^{-5}\,,~~~~~~{\rm at}\,\,\,95\%\,\,{\rm C.L.}\,.
\end{align}
$\bar{M}_{\rm Pl}$ is the reduced Planck mass ($\bar{M}_{\rm Pl}^2 = 1/8\pi G_N \simeq 2.4\times 10^{18}$ GeV).
For us, inflation takes place in a decoupled sector, and we only need to specify $H_{\rm inf}$ and $N_*$. 
We consider a high-scale inflationary model with $H_{\rm inf}/\bar{M}_{\rm Pl} = 10^{-6}$ and we take $N_* = 55$ as the $e$-fold time at which the pivot scale $k_* = 0.05$ Mpc$^{-1}$ exits the Hubble horizon during 
inflation.

The comoving scale $k$ re-enters the horizon after inflation at the temperature 
\begin{align}
T_k = \left(\frac{90 \bar{M}_{\rm Pl}^2 H_{\rm inf}^2}{\pi^2 g_*}\right)^{1/4}e^{-N_*}\left(
\frac{k}{k_*}
\right)\,,
\end{align} 
with $g_*$ the relativistic degrees of freedom.
It is also interesting to mention here that, if we convert the comoving wavenumber into frequency by means of $k=2\pi f$, one finds (see e.g.\,\cite{Franciolini:2022tfm})
\begin{align}
T_{f}\,
\simeq 
61 \, {\rm MeV}\, 
\left (\frac{g_*}{10.75} \right )^{1/6}\left(
\frac{f}{10^{-9} {\rm Hz}}
\right).
\end{align} 
The above estimate implies that comoving wavenumbers corresponding to frequencies of order $\mathcal{O}(10^{-9})$ Hz re-enter the Hubble horizon when the temperature is of the order of the QCD phase transition (see fig.\,\ref{fig:TaleOfScales}).
This is expected to modify any cosmological SGWB in the range of frequency detectable by PTA experiments\,\cite{
Schettler_2011,
Saikawa:2018rcs,
Kohri:2018awv,
Hajkarim:2019nbx,
Abe:2020sqb,
Brandenburg:2021tmp,
inprep_2}. 
Implications of this idea are discussed in sec \ref{Sec:SGWB}.
 \end{itemize}

\subsection{The dynamics after inflation}
\label{sec:AfterInfla}
The angular field $\vartheta$ constitutes a sub-dominant component of the energy-density budget during inflation. 
In order to make the perturbations $\delta\vartheta_k$ phenomenologically relevant, we need to transfer them into radiation after the end of inflation. 
For this to happen, we shall introduce -- following the spirit of conventional curvaton models -- a coupling between the field $\vartheta$ and photons.
After the end of inflation, the inflaton energy density is converted into radiation. 
We assume instantaneous reheating.  The reheating temperature is then given by
\begin{align}\label{eq:Reheat}
T_{\rm RH} = \left(\frac{90}{\pi^2 g_*}\right)^{1/4}\bar{M}_{\rm Pl}^{1/2}H_{\rm inf}^{1/2}
\simeq 10^{15}
{\rm GeV}
\left(\frac{106.75}{g_*}\right)^{1/4}\left(\frac{H_{\rm inf}}{10^{-6}\,\bar{M}_{\rm Pl}}\right)^{1/2}
\,.
\end{align}
In eq.\,(\ref{eq:Reheat}) we take $g_*=  106.75$ (but we remind that $g_*$ is a function of $T$).
We assume that the global $U(1)$ symmetry is broken explicitly by some non-perturbative effect and  the curvaton obtains the following potential
\begin{align}\label{eq:CurvatonMass}
\mathcal{V}(\phi) = \Lambda^4\left(1 - \cos\frac{\phi}{f}\right) \simeq \frac{1}{2}m_{\phi}^2\phi^2\,,
\end{align}
where the curvaton mass is $m_{\phi} = \Lambda^2/f$. 
We now have $\vartheta = \phi/f$. We simplify the analysis and consider in eq.\,(\ref{eq:CurvatonMass})
the quadratic approximation\footnote{If we go beyond the quadratic potential, including for instance anharmonic corrections caused by curvaton self interactions, the equation of motion of the curvaton becomes non-linear and NG acquires a scale-dependence \,\cite{Enqvist_2010}. We leave this case for a future work.}  in which we only have a mass term for $\phi$. 

The curvaton field remains approximately constant until  the Hubble parameter falls below the  curvaton mass when the  curvaton  starts oscillating  around  the  minimum of its potential. 
Notice that the potential in eq.\,(\ref{eq:CurvatonMass}) is generated only after the temperature $T$ of 
the radiation bath drops below some value of the order of the confinement scale $\Lambda$. For instance, in the case of the QCD axion we have
\begin{align}
m_{\phi}^2(T) =
\left\{
\begin{array}{ccc}
 c_0(\Lambda_{\rm QCD}^4/f^2) & {\rm for} & T \lesssim T_0    \\
 c_T(\Lambda_{\rm QCD}^4/f^2)\left(\Lambda_{\rm QCD}/T\right)^{n}  & {\rm for} & T \gtrsim T_0    
\end{array}
\right. ,
\end{align}
with the  parameters $c_0$, $c_T$ and $n$ that can  be determined  using  the dilute 
instanton gas approximation valid at high temperatures. 
Typical values are $c_0 \simeq 10^{-3}$, $c_T \simeq 10^{-7}$ and $n\simeq 7$ with $T_0 \simeq 100$ MeV and $\Lambda_{\rm QCD}\simeq 400$ MeV. 
In principle, we can postulate a situation that mimics the one of QCD, and consider  the temperature-dependent  curvaton mass
\begin{align}\label{eq:ThermalALPmass}
m_{\phi}^2(T) =
\left\{
\begin{array}{ccc}
 m_{\phi}^2 & {\rm for} & T \lesssim \Lambda    \\
 m_{\phi}^2\left(\Lambda/T\right)^{n}  & {\rm for} & T \gtrsim \Lambda    
\end{array}
\right. ,
\end{align}
with zero-temperature value $m_{\phi} = \Lambda^2/f$. 
We  remark that  the  temperature  that controls the curvaton mass does not need to equal the temperature of the SM bath. 
For instance, this is the case if the curvaton mass is generated by couplings to a hidden sector that is not in kinetic equilibrium with the SM.  
For simplicity, we take $T$ in eq.\,(\ref{eq:ThermalALPmass}) to be the temperature of the SM bath (see ref.\,\cite{Blinov:2019rhb} for a critical discussion). 
For definiteness, we set $n=8$ in eq.\,(\ref{eq:ThermalALPmass}).  

During curvaton oscillations, $\rho_{\phi} \propto a^{-3}$ and $\rho_{\gamma} \propto a^{-4}$.
Therefore,  the  curvaton  component  of the energy density $\rho_{\phi}$ increases with respect to the radiation component $\rho_{\gamma}$.
This stage of the dynamics lasts until the Hubble rate $H$ becomes of the order of the decay width $\Gamma_{\phi}$ of the curvaton. After this time, the decay of the curvaton into radiation becomes quickly efficient. 
Schematically, the dynamics after the end of inflation is summarized in the following sketch (with $N = \int Hdt$)
{\small
\begin{align}\label{eq:Sketch}
	\begin{tikzpicture}
	 {\scalebox{1}{
    \draw[->,>=Latex][thick] (-7,0)--(7.,0);
    \draw[thick][thick] (-7,0.1)--(-7,-0.1);
    \node at (-7,-0.4) {\scalebox{1}{$N = 0$}};
    \node at (-7,+0.4) {\scalebox{0.8}{reheating}};
    \node at (-5,+0.2) {\scalebox{0.8}{curvaton constant}};
    \node at (-5,-0.3) {\scalebox{0.8}{{\color{indiagreen}{``first'' radiation stage}}}};
    \node at (-5,+0.75) {\scalebox{0.9}{{\color{cornellred}{Phase I}}}};
    \draw[thick][thick] (-3,0.1)--(-3,-0.1);
    \node at (-3,+0.4) {\scalebox{1}{$H \sim m_{\phi}$}};
    \node at (-3,-0.4) {\scalebox{1}{$N_{\rm osc}$}};
    \node at (-0.5,+0.2) {\scalebox{0.8}{curvaton oscillates}};
    \node at (-0.5,+0.75) {\scalebox{0.9}{{\color{cornellred}{Phase II}}}};
    \node at (-0.5,-0.25) {\scalebox{0.8}{$P_{\phi}\simeq 0$, $\rho_{\phi}\sim a^{-3}$ and $\rho_{\gamma}\sim a^{-4}$}};
    \draw[thick][thick] (2.,0.1)--(2.,-0.1);
    \node at (2.,+0.4) {\scalebox{1}{$H \sim \Gamma_{\phi}$}};
    \node at (2.,-0.4) {\scalebox{1}{$N_{\rm dec}$}};
    \node at (4.5,+0.75) {\scalebox{0.9}{{\color{cornellred}{Phase III}}}};    
    \node at (4.5,+0.2) {\scalebox{0.8}{curvaton decays to radiation}};
    \node at (4.5,-0.3) {\scalebox{0.8}{{\color{indiagreen}{``second'' radiation stage}}}};
    \node at (-5,1.2) {\scalebox{0.8}{{\color{VioletRed4}{\textbf{Schematic of background evolution}}}}};
    }}
	\end{tikzpicture}
\end{align}
}
We assume that $\phi$ decays as consequence of the following dimension-5 effective operator 
\begin{align}
\mathcal{L}_{\phi F\tilde{F}} = \frac{g_{a\gamma\gamma}}{4f}\,\phi F_{\mu\nu}\tilde{F}^{\mu\nu}\,,
\end{align}
which gives the decay rate $\Gamma_{\phi}$ 
\begin{align}\label{eq:Gamma}
\Gamma_{\phi} &= \frac{g_{a\gamma\gamma}^2 m_{\phi}^3}{64\pi f^2} \simeq 
g_{a\gamma\gamma}\left(\frac{\Lambda}{10^4\,{\rm GeV}}\right)^6
\left(\frac{10^{11}\,{\rm GeV}}{f}\right)^5\left(\frac{10^{8}}{\tau_{\rm U}}\right) 
\,,
\end{align}
that we have written in erms of the present age of the Universe $\tau_{\rm U} \simeq 13.8\times 10^{9}$ yr, while the mass $m_{\phi}$ is
\begin{align}
m_{\phi} & \simeq 10^6 \,{\rm eV}
\left(\frac{\Lambda}{10^4\,{\rm GeV}}\right)^2
\left(\frac{10^{11}\,{\rm GeV}}{f}\right)\,.
\end{align}
The lifetime $\tau_{\phi} = 1/\Gamma_{\phi}$ can be either much longer or much shorter than $\tau_{\rm U}$.
Typically, the cosmologically long-lived option $\tau_{\phi} \gg \tau_{\rm U}$ is preferred  
since in this case $\phi$ can naturally serve as dark matter whose abundance 
is generated by means of the misalignment mechanism (see ref.\,\cite{Marsh:2015xka,DiLuzio:2020wdo} for a review), but in the opposite regime, as we are considering in this work, $\phi$ is not cosmologically stable and decays into radiation.
 
In order to describe the dynamics of the model after the end of inflation, we study Einstein and fluid equations.
Consider first the unperturbed background. 
The Friedmann equation and the continuity equation are 
\begin{align}\label{eq:Fluids}
{\rm Friedmann\,equation:}&~~~~3\bar{M}_{\rm Pl}^2 H^2 = \rho_{\gamma} + \rho_{\phi}\,,
\nn
\\
{\rm continuity\,equation\,for\,each\,fluid:}&~~~~\dot{\rho}_{\alpha}  = -3H(\rho_{\alpha} + P_{\alpha}) + Q_{\alpha}\,,
\end{align}
where in this case we have two background fluids, that are the radiation fluid 
with energy density $\rho_{\gamma} = \rho_{\gamma}(t)$ and pressure $P_{\gamma} = P_{\gamma}(t) = \rho_{\gamma}/3$ and the homogeneous field $\phi = \phi(t)$ whose energy density and pressure are 
given by\footnote{The energy-momentum tensor in the FLRW geometry is given by 
$T_{\alpha}^{\mu\nu} = {\rm diag}(\rho_{\alpha},P_{\alpha}/a^2,P_{\alpha}/a^2,P_{\alpha}/a^2)$ 
and for the scalar field $\phi$ energy density and pressure can be identified if one computes the components 
of $T_{\phi}^{\mu\nu} = (\partial^{\mu}\phi)(\partial^{\nu}\phi) - g^{\mu\nu}[(\partial_{\rho}\phi)(\partial^{\rho}\phi)/2 + \mathcal{V}(\phi)]$. 
The continuity equation $\dot{\rho}_{\alpha} = -3H(\rho_{\alpha} + P_{\alpha})$ (in the absence of energy transfer, $\Gamma_{\phi} = 0$) 
simply follows from the temporal component of $T_{\alpha~~;\mu}^{\mu\nu} = 0$. 
In the presence of interactions between the fluids, notice that in general we have $T_{\alpha~~;\mu}^{\mu\nu} \neq 0$ for each individual fluid but, as a consequence of the Bianchi identity, we must have the conservation equation 
$\sum_{\alpha}T_{\alpha~~;\mu}^{\mu\nu} = 0$ for their sum. 
This condition forces the relation $Q_{\gamma} = - Q_{\phi}$. 
We refer to appendix\,\ref{app:Pert} for a more detailed discussion.
}
\begin{align}\label{eq:EnergyAndPressure}
\rho_{\phi} = \frac{1}{2}\dot{\phi}^2 + \mathcal{V}(\phi)\,,~~~~~P_{\phi} = \frac{1}{2}\dot{\phi}^2 - \mathcal{V}(\phi)\,.
\end{align}
In eq.\,(\ref{eq:Fluids}) $Q_{\alpha}$ represents the energy transfer per unit time to the $\alpha$-fluid. 
This transfer of energy is due to the decay $\phi\to \gamma\gamma$ previously introduced, and 
we have $Q_{\gamma} = - Q_{\phi} = \Gamma_{\phi}\rho_{\phi}$. 
All in all, the three relevant equations are 
\begin{align}
3\bar{M}_{\rm Pl}^2 H^2 & = \rho_{\gamma} + \rho_{\phi}\equiv \rho\,,\label{eq:DynBG1}\\
\dot{\rho}_{\gamma} + 4H\rho_{\gamma} & = +\Gamma_{\phi}\rho_{\phi}\,,\label{eq:DynBG2}\\
\dot{\rho}_{\phi} + 3H(\rho_{\phi} + P_{\phi}) & = - \Gamma_{\phi}\rho_{\phi}\,.\label{eq:DynBG3}
\end{align}
During phase II (see eq.\,(\ref{eq:Sketch})) the curvaton oscillates and one can take the 
approximation $P_{\phi} \simeq 0$. 
This is because during this phase the curvaton field oscillates with a typical timescale set by its mass, $t_{\phi} = 1/m_{\phi}$, which is smaller than the inverse Hubble rate, $t_{\phi} < 1/H$.
One can, therefore, average over an oscillation and estimate $\dot{\phi} \simeq m_{\phi}\phi$; 
from eq.\,(\ref{eq:EnergyAndPressure}) it follows that $P_{\phi}\simeq 0$. 
This is the approximation used in ref.\,\cite{Firouzjahi:2012iz}. 
This is a good description of the classical field dynamics when the Hubble rate drops below the mass of the curvaton.\footnote{In the limit $P_{\phi} \to 0$ the continuity equation 
$\dot{\rho}_{\phi} + 3H(\rho_{\phi} + P_{\phi}) = - \Gamma_{\phi}\rho_{\phi}$ can be written in the form $\ddot{\phi} + (3H + \Gamma_{\phi})\dot{\phi} + \mathcal{V}^{\prime}(\phi) = 0$ and the factor $\Gamma_{\phi}$ enters in the Klein-Gordon equation in the form of a damping term.}

\begin{figure}[!h!]
\begin{center}
$$
\includegraphics[width=.33\textwidth]{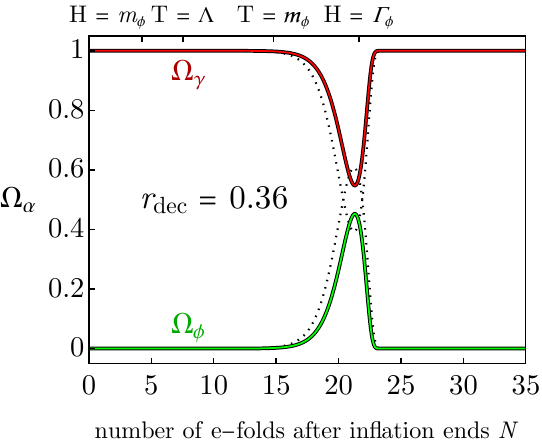}
\includegraphics[width=.33\textwidth]{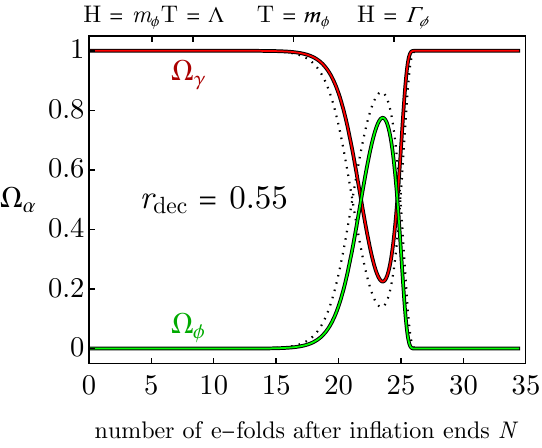}
\includegraphics[width=.33\textwidth]{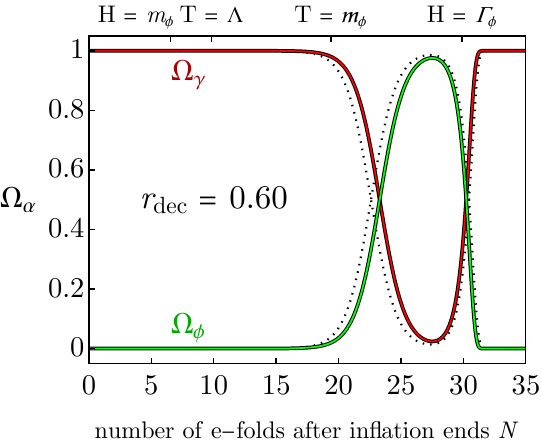}$$
\caption{\em \label{fig:EvoAfterInfla} 
Here we set $f = 10^{15}$ GeV, $H_{\rm inf}/\bar{M}_{\rm Pl} = 10^{-6}$, $\vartheta_0 = 0.01$. 
In the left panel we have $m_{\phi} = 5\times10^8$ GeV, in the central panel $m_{\phi} = 10^8$ GeV and 
in the right panel $m_{\phi} = 5\times 10^7$ GeV.
The $e$-fold time difference $\Delta N$ between $T=m_{\phi}$ and $H=\Gamma_{\phi}$ increases from left to right.
This implies that the curvaton energy density has progressively more time to increase  with respect to radiation (since during this time interval it redshifts as non-relativistic matter) before decaying into the latter. 
In each panel we quote the corresponding value of the parameter $r_{\rm dec}$ defined in eq.\,(\ref{eq:rDef}). 
The numerical value of $r_{\rm dec}$ is computed at the time $N_{\rm dec}$ at which $H = \Gamma_{\phi}$, as indicated on the labels of the top $x$-axis.
Dashed lines represent the model in the simplified setup (further details are provided in the main text). } 
\end{center}
\end{figure}

We solve the system in eqs.\,(\ref{eq:DynBG1}-\ref{eq:DynBG3}) in two steps. 
First, we consider the evolution during phase I. 
We set the initial conditions $\rho_{\gamma}(N=0) = 3\bar{M}_{\rm Pl}^{2}H_{\rm inf}^{2}$ 
and $\vartheta(N=0) = \vartheta_0$ (with vanishing initial velocity). We rewrite eq.\,(\ref{eq:DynBG3}) in terms of an evolution equation for $\vartheta(N)$ with energy density and pressure given by eq.\,(\ref{eq:EnergyAndPressure}), and we use the number of $e$-folds as time variable. 
To be more precise, the system we solve is
\begin{align}
 \frac{dH}{dN} + \frac{3H}{2} + \frac{H\rho_{\gamma}}{2\rho} + 
\frac{3HP_{\phi}}{2\rho} & = 0\,, \label{eq:DynBGSim1}
\\
\frac{d\rho_{\gamma}}{dN} + 4\rho_{\gamma} - \frac{f^2\Gamma_{\phi}H}{2}\bigg[
\left(\frac{d\vartheta}{dN}\right)^2 + 
\frac{m_{\phi}(T)^2}{H^2}\vartheta^2
\bigg] &= 0\,,
\\
\frac{d\vartheta}{dN}\frac{d^2\vartheta}{dN^2} 
+ \bigg(\frac{d\vartheta}{dN}\bigg)^2\bigg[3+\frac{d}{dN}\log H\bigg] 
+
\frac{m_{\phi}(T)^2}{H^2}\vartheta\bigg[\frac{d\vartheta}{dN} 
-\frac{n\vartheta}{2}\frac{d}{dN}\log T\bigg]&
\nn \\
+
\frac{\Gamma_{\phi}}{2H}\left[
\left(\frac{d\vartheta}{dN}\right)^2 + 
\frac{m_{\phi}(T)^2}{H^2}\vartheta^2
\right] & = 0\,. \label{eq:DynBGSim3}
\end{align}
Notice that, as in the case of the conventional misalignment mechanism when the global $U(1)$ symmetry gets spontaneously broken during inflation, $\vartheta_0$ is a free parameter.
We follow the evolution of the system until the field $\vartheta$ and its pressure $P_{\phi}$ start oscillating on timescales shorter than the inverse Hubble rate. At this point, as discussed before, the pressure can be safely neglected. Numerically, this first part of the dynamics lasts until $T = \Lambda$ at $e$-fold time $N_{\Lambda}$.\footnote{Numerically, we checked that the exact value of $N$ at which we switch to phase II does not impact our results as long as $N \gtrsim N_{\Lambda}$.}
The subsequent evolution can be described by the same system in eqs.\,(\ref{eq:DynBG1}-\ref{eq:DynBG3}) in which we now set $P_{\phi} = 0$.  
In this limit the system simplifies to 
\begin{align}
\frac{dH}{dN} + \frac{3H}{2} + \frac{H\Omega_{\gamma}}{2} & = 0\,,\label{eq:DynSim1}\\
\frac{d\Omega_{\gamma}}{dN} + \Omega_{\phi}\Omega_{\gamma} - \frac{\Gamma_{\phi}\Omega_{\phi}}{H} & = 0\,,\\
\frac{d\Omega_{\phi}}{dN} - \Omega_{\phi}\Omega_{\gamma} + \frac{\Gamma_{\phi}\Omega_{\phi}}{H} & = 0\,,\label{eq:DynSim3}
\end{align}
where we introduce the time-dependent quantities $\Omega_{\gamma} \equiv \rho_{\gamma}/\rho$ and 
$\Omega_{\phi} \equiv \rho_{\phi}/\rho$ (with $\Omega_{\gamma}+\Omega_{\phi} = 1$ by definition). 
We solve the system in eqs.\,(\ref{eq:DynSim1}-\ref{eq:DynSim3}) with initial conditions given by the solutions of eqs.\,(\ref{eq:DynBGSim1}-\ref{eq:DynBGSim3}) at time $N = N_{\Lambda}$.
We show in fig.\,\ref{fig:EvoAfterInfla} the background dynamics in terms of $\Omega_{\gamma}$ and $\Omega_{\phi}$ after the end of inflation for three benchmark sets of parameters. 

It is also possible to describe the dynamics in a simplified way. 
During phase I, we can set $\Gamma_{\phi} = 0$ and $\rho_{\phi} = 0$ (massless field frozen at some initial constant value). Furthermore, we neglect the time-dependence of $g_*$ and set $g_* = 106.75$. This is correct as long as this part of the dynamics takes palace before the electroweak phase transition.  
Under these assumptions, we find the following analytical solutions for the temperature and the Hubble rate
\begin{align}
T(N) = \left(\frac{90 \bar{M}_{\rm Pl}^2 H_{\rm inf}^2}{\pi^2 g_*}\right)^{1/4}e^{-N}\,,~~~~~~~~H(N) = H_{\rm inf}e^{-2N}\,.
\end{align}
In this simplified setup, we can compute the $e$-fold time $N_{\Lambda}$ defined by $T(N_{\Lambda}) \equiv \Lambda$ and 
the $e$-fold time $N_{\phi}$ defined by $H(N_{\phi}) \equiv m_{\phi}$. We find
\begin{align}
N_{\Lambda} = 
\frac{1}{4}\log\left(\frac{90\bar{M}_{\rm Pl}^2 H_{\rm inf}^2}
{\pi^2g_*\Lambda^4}\right)\,,~~~~~~~
N_{\phi} = \frac{1}{2}\log\left(\frac{H_{\rm inf}}{m_{\phi}}\right)\,.
\end{align}
Notice that if we consider the case $f \ll \bar{M}_{\rm Pl}$ then the previous relations (we remind that $m_{\phi} = \Lambda^2/f$) imply that $N_{\phi} < N_{\Lambda}$ (see fig.\,\ref{fig:EvoAfterInfla}). 
This means that at time $N_{\phi}$ when $H(N_{\phi}) = m_{\phi}$ the temperature is of order $T(N_{\phi}) \simeq (\bar{M}_{\rm Pl}/f)^{1/2}\Lambda \gg \Lambda$ and the field $\vartheta$ does not fill yet the effect of the non-zero mass. This means that identifying the time $N_{\rm osc}$ with $N_{\phi}$ would not be entirely correct. 
On the other hand, at time $N_{\Lambda}$ the temperature is $T(N_{\Lambda}) = \Lambda$ and the mass $m_{\phi}(T)$ fully formed (see eq.\,(\ref{eq:ThermalALPmass})). 
It is, therefore, more realistic to identify $N_{\rm osc}$ with $N_{\Lambda}$.
At this time, we estimate the initial condition for the energy density $\Omega_{\phi}$ to be  $\Omega_{\phi}(N_{\rm osc}) = \Lambda^4\vartheta_0^2/3\bar{M}_{\rm Pl}^2H(N_{\rm osc})^2$ while  for the radiation bath we have $\Omega_{\gamma}(N_{\rm osc}) = 1 - \Omega_{\phi}(N_{\rm osc})$. Given these initial conditions, we solve the system in eqs.\,(\ref{eq:DynSim1}-\ref{eq:DynSim3}) for $N > N_{\rm osc}$. 
The result is also shown in fig.\,\ref{fig:EvoAfterInfla} (dotted black lines).

Notice that it is important to have good control over the initial value of $\Omega_{\phi}$ that is used to solve the system in eqs.\,(\ref{eq:DynSim1}-\ref{eq:DynSim3}) since this value gets exponentially modified during the dynamical evolution in phase II. 
In the so-called sudden decay limit (that is $\Gamma_{\phi} = 0$) we indeed have the analytical solutions\,\cite{Firouzjahi:2012iz}
\begin{align}\label{eq:SuddenDecay}
\Omega_{\gamma}(N) = \frac{1}{1+pe^{N}}\,,~~~~~~~\Omega_{\phi}(N) = \frac{pe^{N}}{1+pe^{N}}\,,
\end{align}
with $p\equiv \Omega_{\phi}/\Omega_{\gamma}$ at $N = N_{\rm osc}$. 

\begin{figure}[h]
\begin{center}
$$\includegraphics[width=.445\textwidth]{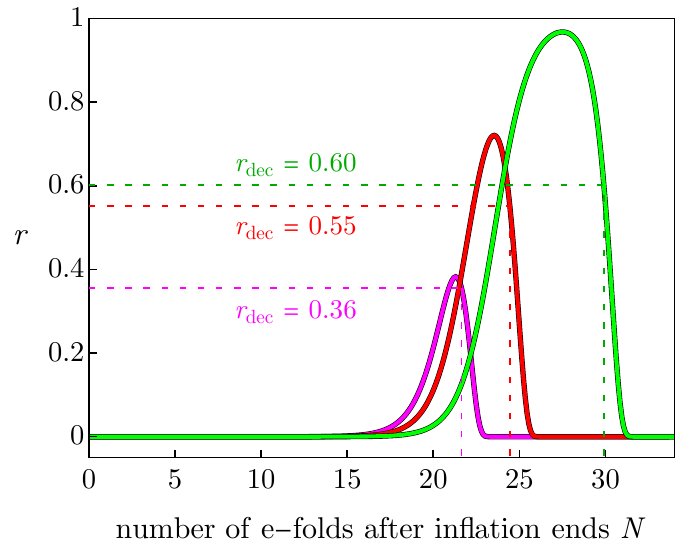}
\qquad\includegraphics[width=.45\textwidth]{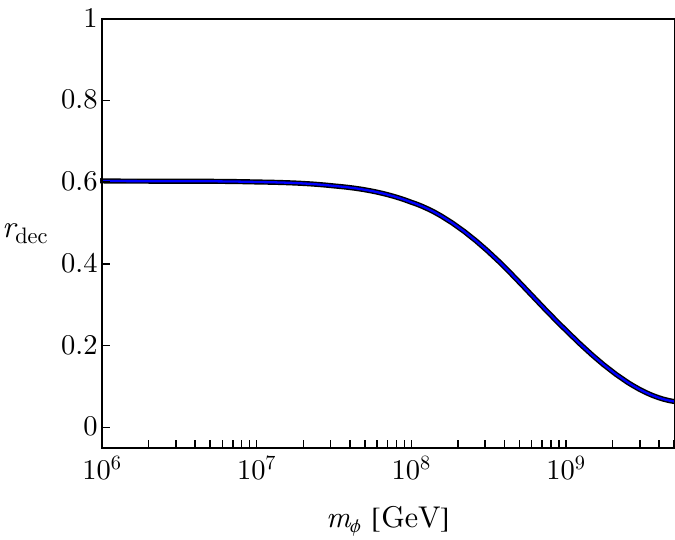}$$
\caption{\em
We fix 
$f = 10^{15}$ GeV, $H_{\rm inf}/\bar{M}_{\rm Pl} = 10^{-6}$, $\vartheta_0 = 0.01$.  
In the left panel, we show the evolution of $r$ for the three values of $m_{\phi}$ shown in fig.\,\ref{fig:EvoAfterInfla}; we also 
identify the value of $r_{\rm dec}$, defined as in eq.\,(\ref{eq:rDef}). 
In the right panel, we show the value of $r_{\rm dec}$ as function of $m_{\phi}$.
 }\label{fig:rDec}  
\end{center}
\end{figure}

If we define the time-dependent quantity $r(N) \equiv 3\rho_{\phi}/(3\rho_{\phi} + 4\rho_{\gamma})$, a key parameter in the analysis of the following sections will be the weighted fraction of the curvaton energy density to the total energy density at the time of curvaton decay, defined by\footnote{Notice that in refs.\,\cite{Ando:2017veq,Inomata:2020xad} this parameter is differently defined as the ratio between the energy densities of the curvaton and radiation. More concretely, refs.\,\cite{Ando:2017veq,Inomata:2020xad} define $r$ to be $\kappa \equiv \rho_{\phi}/\rho_{\gamma}$, and the relation with $r$ used in this work is given by $r = 3\kappa/(4+3\kappa)$. 
We, therefore, adopt the more conventional definition of $r \equiv 3\rho_{\phi}/(3\rho_{\phi} + 4\rho_{\gamma})$ used in the curvaton literature.}
\begin{align}\label{eq:rDef}
r_{\rm dec} \equiv r(N_{\rm dec}) = 
\left.
\frac{3\rho_{\phi}}{
3\rho_{\phi} + 4\rho_{\gamma}
}\right|_{N = N_{\rm dec}}
= \left.\frac{3\Omega_{\phi}}{4 - \Omega_{\phi}}\right|_{N = N_{\rm dec}}\,,~~~~~
{\rm with}\,\,\,H = \Gamma_{\phi}\,\,\,{\rm at\,\,}N=N_{\rm dec}\,.
\end{align}
We show the evolution of the parameter $r$ in fig.\,\ref{fig:rDec}. 
More in detail, the left panel of fig.\,\ref{fig:rDec} shows the evolution of $r$ corresponding to the three benchmark models discussed in fig.\,\ref{fig:EvoAfterInfla}; the right-hand panel, on the contrary, shows the value of $r_{\rm dec}$ as function of $m_{\phi}$ (with the parameters $H_{\rm inf}$, $f$ and $\vartheta_0$ fixed to the values quoted in the caption and in fig.\,\ref{fig:EvoAfterInfla}).
The time evolution of $r$ clearly retraces the time evolution of $\Omega_{\phi}$, and features a maximum during the oscillating phase of the curvaton.
Notice that the numerical value of $r_{\rm dec}$ depends on its specific definition. 
If we define $r_{\rm dec}$ strictly as in eq.\,(\ref{eq:rDef}), then we find $r_{\rm dec} \lesssim 0.6$ even for cases in which the energy density of the curvaton happens to dominate the total energy density of the Universe before decaying (see the plateau in the right panel of fig.\,\ref{fig:rDec}). This is due to the fact that the decay of the curvaton is not an instantaneous process that happens at $H = \Gamma_{\phi}$ but, as evident from fig.\,\ref{fig:EvoAfterInfla} and fig.\,\ref{fig:rDec}, when $H = \Gamma_{\phi}$ part of the curvaton energy density was already converted back into radiation.

The importance of the parameter $r_{\rm dec}$ is twofold. On the one hand, at the linear level, 
it controls the fraction of perturbations that are transferred to radiation;
we shall discuss in more detail this effect in section\,\ref{eq:Per}.  
On the other hand, it also controls the impact of non-gaussian corrections; we shall discuss this issue and its implications in section\,\ref{sec:Pheno}.

At this stage of the analysis, we can pause to try a first exploration of the parameter space.
Cosmologically, the relevant parameters are $N_{\rm dec}$ and $r_{\rm dec}$. 
As far as the fundamental parameters are concerned, we focus our attention primarily on $m_{\phi}$ and $f$.
\begin{figure}[!h!]
\begin{center}
$$\includegraphics[width=.46\textwidth]{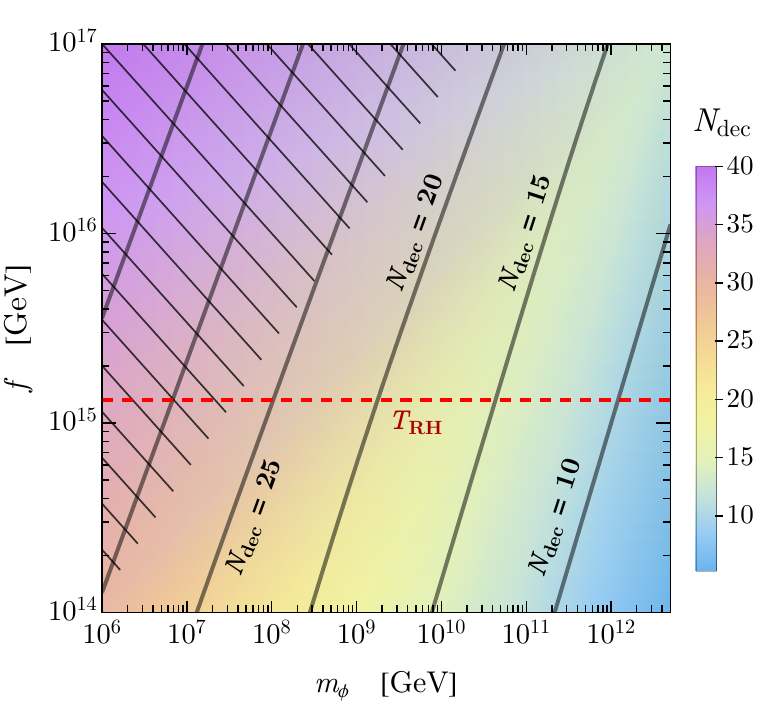}
\qquad\includegraphics[width=.46\textwidth]{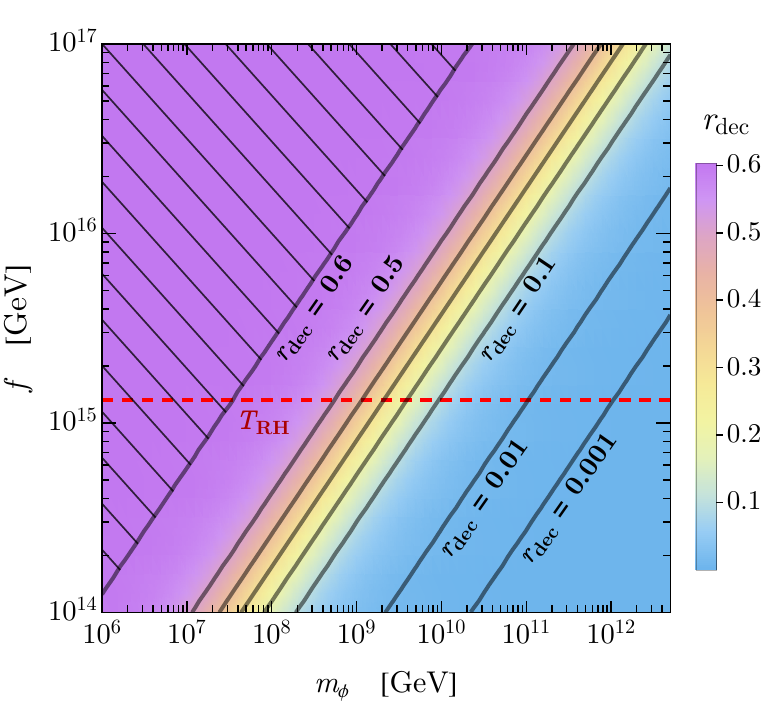}$$
\caption{\em 
$\vartheta_0 = 0.01$.  
Scan over the parameter space of the model in light of the background analysis discussed in section\,\ref{sec:AfterInfla}.  
The horizontal dashed red line indicates the assumed re-heating temperature scale \eqref{eq:Reheat}. The black hatched region identifies the parameter space for which $\tau_{\phi} \geq \tau_{\rm U}$.
 }\label{fig:MiniScan}  
\end{center}
\end{figure}
In fig.\,\ref{fig:MiniScan} we fix $H_{\rm inf}/\bar{M}_{\rm Pl} = 10^{-6}$ and, for the moment, we keep $\vartheta_0 = 0.01$.

\begin{figure}[!h!]
\begin{center}
$$\includegraphics[width=.45\textwidth]{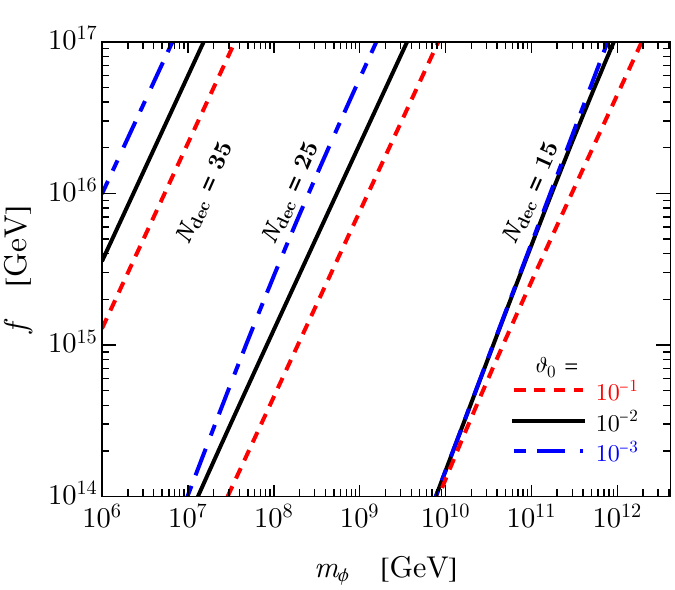}
\qquad\includegraphics[width=.45\textwidth]{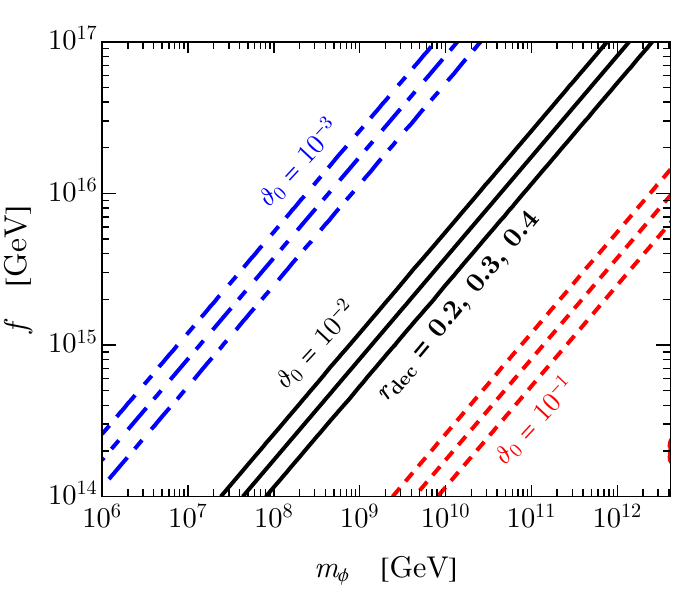}$$
\caption{\em 
Same as in fig.\,\ref{fig:MiniScan} but for different values of $\vartheta_0$ 
($\vartheta_0 = 10^{-1}$, dashed red lines; $\vartheta_0 = 10^{-3}$, dot-dashed blue lines; solid black lines 
refer to $\vartheta_0 = 10^{-2}$). As far as $N_{\rm dec}$ is concerned, on the left panel we only show contours 
corresponding to $N_{\rm dec} = 15,25,35$; similarly, on the right panel, we only show contours 
corresponding to $r_{\rm dec} = 0.2,0.3,0.4$ (from the right- to the left-most lines in each of the three groups).
 }\label{fig:MiniScanTheta}  
\end{center}
\end{figure}

We comment about the $\vartheta_0$-dependence in fig.\,\ref{fig:MiniScanTheta}. 
Intuitively, we do not expect a strong $\vartheta_0$-dependence for what concerns $N_{\rm dec}$ since 
the latter is determined by the equation $H = \Gamma_{\phi}$. In this equation, the decay width $\Gamma_{\phi}$ 
does not depend on $\vartheta_0$; the value of the Hubble parameter, on the contrary, does 
depend on $\vartheta_0$ since the size of $\vartheta_0$ controls the (initial value of the) curvaton energy density 
which enters in the Friedmann equation, eq.\,(\ref{eq:DynBG1}).  
However, as long as the curvaton does not dominate the energy budget of the Universe for long time, this change 
does not alter much the evolution of $H$ and, consequently, the time at which $H = \Gamma_{\phi}$.  
This is confirmed by the numerical analysis shown in the left panel of fig.\,\ref{fig:MiniScanTheta} (see caption for details). 
Consider now the value of $r_{\rm dec}$. Contrary to the previous argument, we do expect a sizable dependence on $\vartheta_0$.
The reason follows from what we already noticed above eq.\,(\ref{eq:SuddenDecay}):
the value of the curvaton energy density at the beginning of the oscillating phase (controlled by $\vartheta_0$) 
gets exponentially modified by the dynamics during Phase II.

\subsection{Perturbations at linear order: curvature power spectrum }\label{eq:Per}

At the linear order, the gauge invariant curvature perturbation on spatial slices of uniform energy
density reads\,\cite{Kodama:1984ziu}
\begin{equation}
\zeta \equiv -\psi - H\frac{\delta\rho}{\dot{\rho}}\,,
\end{equation}
with $\rho = \rho_{\phi} + \rho_{\gamma}$, 
$\dot{\rho} = \dot{\rho}_{\phi} + \dot{\rho}_{\gamma}$
 and $\delta\rho = \delta\rho_{\phi} + \delta\rho_{\gamma}$; 
 $\psi$ is the gauge-dependent curvature perturbation that enters in the 
 linear perturbations about a spatially-flat Friedmann-Robertson-Walker background, as defined in eq.\,(\ref{eq:MetricPerturbation}).
 We refer to appendix\,\ref{app:Pert} for a more detailed discussion.

It is customary to introduce individual curvature perturbations $\zeta_{\alpha}$ each of which is associated with a single energy density component, and 
similarly defined by 
$\zeta_{\alpha} \equiv -\psi - H(\delta\rho_{\alpha}/\dot{\rho}_{\alpha})$. 
Consequently, the curvature perturbation $\zeta$ can be equivalently written as  
\begin{equation}\label{eq:TotalZeta1}
\zeta = -\psi + \frac{[\dot{\rho}_{\phi}(\zeta_{\phi} + \psi) + 
\dot{\rho}_{\gamma}(\zeta_{\gamma} + \psi)
]}{\dot{\rho}} = \frac{\dot{\rho}_{\phi}}{\dot{\rho}}\,\zeta_{\phi} + 
\frac{\dot{\rho}_{\gamma}}{\dot{\rho}}\,\zeta_{\gamma}\,.
\end{equation}
This equation is exact.  
We are interested in the evolution of $\zeta$ in Fourier space, and more concretely in its power spectrum.  
In other words, the goal is computing the quantity 
\begin{align}\label{eq:FinaalPS}
P_{\zeta}(k) = \frac{k^3}{2\pi^2}|\zeta_k(N_{\rm f})|^2\,,
\end{align}
where $\zeta_k(N)$ is the time-dependent Fourier mode of $\zeta$ for a given comoving wavenumber $k \equiv |\vec{k}|$ 
and the notation in eq.\,(\ref{eq:FinaalPS}) means that $\zeta_k(N)$ should be evaluated at
some appropriate time $N_{\rm f}$ after the mode $\zeta_k$ settles to its final value, that is conserved until horizon re-entry.  
Importantly, the dynamics of $\zeta$ in controlled by the equation
\begin{align}\label{eq:KeyDynamics}
\frac{d\zeta_k}{dN} = -\frac{\delta P_{{\rm nad},k}}{\rho + P} + \frac{k^2}{3(aH)^2}\big(\Psi_k - \mathcal{R}_k\big)\,,
\end{align}
where $\mathcal{R}$ is the total comoving curvature perturbation, $\Psi$ the curvature perturbation on uniform shear hypersurfaces and $\delta P_{\rm nad}$ the total non-adiabatic pressure perturbation 
(see appendix\,\ref{app:Pert}). This equation implies that $\zeta_k$ is conserved on super-horizon scales (that is when $k\ll aH$ and the last term in eq.\,(\ref{eq:KeyDynamics}) can be safely neglected) and in the absence of non-adiabatic pressure perturbations (that is the first term on the right-hand side of eq.\,(\ref{eq:KeyDynamics})).
Since during phase II+III we have a non-zero $\delta P_{\rm nad}$ (because of the relative entropy perturbation between the curvaton and the radiation fluid, see eq.\,(\ref{eq:Pnad})), the time $N_{\rm f}$ is given by {\it i)} anytime in between $N_{\rm dec} < N_{\rm f} < N_{k}$ if the mode $k$ re-enters the horizon (at time $N_k$ such that $k/aH = 1$) after the curvaton decay takes place or {\it ii)} simply by $N_{\rm f} \simeq N_{k} < N_{\rm dec}$ if the mode $k$ re-enters the horizon before the curvaton decay.
\begin{figure}[!h!]
\begin{center}
$$\includegraphics[width=.45\textwidth]{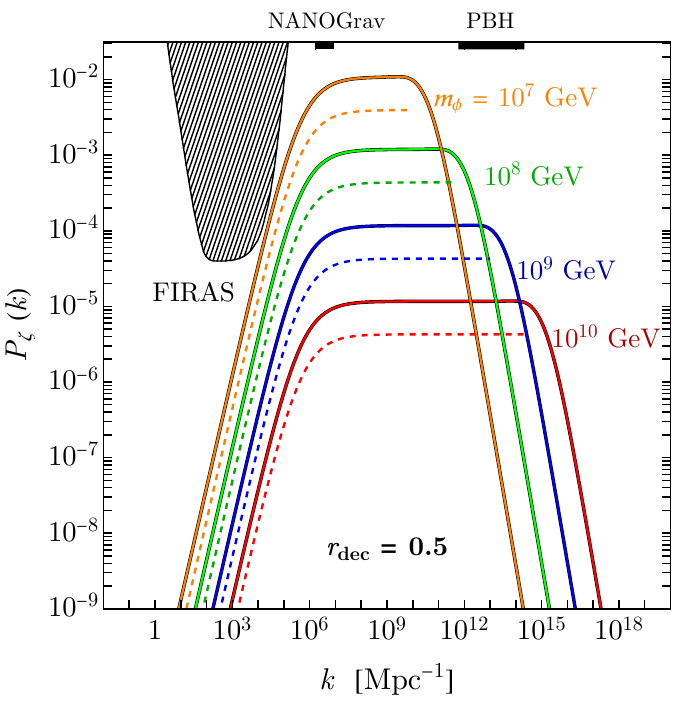}
\qquad\includegraphics[width=.45\textwidth]{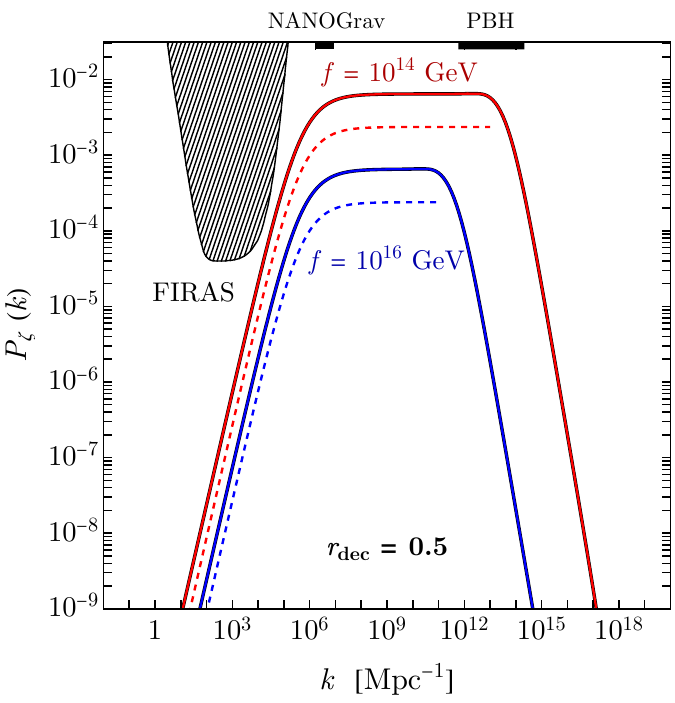}$$
\caption{\em 
Left panel. Power spectra $P_{\zeta}(k)$ in eq.\,(\ref{eq:FinaalPS}) 
for four different representative values of $m_{\phi}$. 
We fix $f = 3\times 10^{15}$ GeV and tune $\vartheta_0$ to get $r_{\rm dec} = 0.5$ 
in each of the four spectra. The dashed lines correspond to the sudden decay approximation while the solid lines 
include the dynamics of phase II+III (see the schematic in eq.\,(\ref{eq:Sketch}) and discussion in section\,\ref{eq:Per}).
Right panel. Same as in the left panel but with $m_{\phi} = 10^8$ GeV fixed and for two representative values of 
the decay constant $f$. 
To guide the eye, on the top x-axis we indicate the $k$-range that is relevant for 
the generation of second-order GWs detectable by the NANOGrav and other PTA experiments and 
the $k$-range that is viable for the identification of the totality of dark matter with PBHs. The dashed regions shows the experimental constraints coming from the
analysis of CMB spectral distortion by the FIRAS collaboration\,\cite{Fixsen:1996nj,Chluba:2012gq,Chluba:2012we,Bianchini:2022dqh}. }\label{fig:PS}  
\end{center}
\end{figure}

Schematically, we describe the dynamics of perturbations for a given $k$-mode as summarized in the following sketch:
{\small
\begin{align}\label{eq:Sketch2}
	\begin{tikzpicture}
	 {\scalebox{1}{
    \draw[->,>=Latex][thick] (-8.5,0)--(6.2,0);
    \draw[thick][thick] (-5,0.1)--(-5,-0.1);
    \node at (-5,-0.4) {\scalebox{1}{$N = 0$}};
    \node at (-6.75,+0.75) {\scalebox{0.9}{{\color{cornellred}{Inflation}}}};
    \node at (-7.1,-0.3) {\scalebox{0.9}{dynamics described}};
    \node at (-6.85-0.5,-0.7) {\scalebox{0.9}{in terms of $\delta\vartheta_k$}};
    \node at (-6.8,+0.3) {\scalebox{0.9}{{\color{cornellred}{$\varphi$ dynamics boosts $\delta\vartheta_k$}}}};
    \node at (-2-0.6,+0.75) {\scalebox{0.9}{{\color{cornellred}{Phase I}}}};
    \node at (-2.6,+0.3) {\scalebox{0.9}{\color{cornellred}{No curvaton decay, $\Gamma_{\phi} = 0$}}}; 
    \node at (-2.9,-0.3) {\scalebox{0.9}{dynamics of $\delta\vartheta_k$}}; 
    \node at (-2.35,-0.7) {\scalebox{0.9}{bridges $N=0$ and $N_{\rm osc}$}}; 
    \node at (-2.85,-1.1) {\scalebox{0.9}{eqs.\,(\ref{eq:Zetaaxion},\,\ref{eq:Raxion})}};  
    \draw[thick][thick] (-0.3,0.1)--(-0.3,-0.1); 
    \node at (-0.3,-0.4) {\scalebox{1}{$N_{\rm osc}$}};
    \node at (2.4,-0.3) {\scalebox{0.9}{dynamics described in terms of}};
    \node at (3.15-0.5,-0.7) {\scalebox{0.9}{$\zeta_k$, $\zeta_{\phi,k}$, $\mathcal{R}_k$ and $\mathcal{R}_{\phi,k}$ with $\Gamma_{\phi} \neq 0$}};
    \node at (2.2,-1.1) {\scalebox{0.9}{eqs.\,(\ref{eq:Final1},\,\ref{eq:Final2},\,\ref{eq:Final3},\,\ref{eq:Final4})}};
    \node at (2.65,+0.75) {\scalebox{0.9}{{\color{cornellred}{Phase II + Phase III}}}};    
    \node at (2.65,+0.3) {\scalebox{0.9}{{\color{cornellred}{curvaton oscillates and decays to radiation}}}};
    \draw[thick][thick] (5.6,0.1)--(5.6,-0.1);        
    \node at (5.6,-0.4) {\scalebox{1}{$N_{\rm f}$}};    
    \node at (-5.4,1.2) {\scalebox{0.8}{{\color{VioletRed4}{\textbf{Dynamics of $k$-perturbation, 
    from $\delta\vartheta_k$ to $\zeta_k$}}}}};
    }}
	\end{tikzpicture}
\end{align}
}

Fig.\,\ref{fig:PS} shows $P_\zeta(k)$, computed by choosing the proper $N_{\rm f}$ for each $k$, for different values of $f$ and $m_\phi$. 
We can deduce the qualitative behaviour of the power spectrum shape under a change of parameters. We observe that decreasing $f$ has the effect of increasing the amplitude and enlarging the range of $k$ where $P_{\zeta}(k)$ is enhanced. This can be explained by considering that the fast-rolling dynamics of $\varphi$ lasts longer for smaller values of $f$ (see also eq.\,(\ref{eq:Gamma})). Hence, angular perturbations have more time to grow (see also eq.\,(\ref{eq:PowerSpectrumRange}). Moreover, as the curvaton decay width scales as $\Gamma_{\phi}\propto f^{-2}$, decreasing $f$ means anticipating decay into radiation. An early-time decay interests larger scales, resulting in a broader power spectrum. Analogously, decreasing $m_{\phi}$ shrinks the power spectrum and pushes it up to higher amplitude values. The reason to this has to be found, again, in the scaling of the decay width $\Gamma_{\phi}\propto m_{\phi}^3$. Also, decreasing the curvaton mass has the effect of increasing the $e$-fold time difference $\Delta N$ between $T = m_\phi$ and $H = \Gamma_\phi$ implying that now the angular perturbations, and hence the amplitude of the power spectrum, have more time to grow (see also fig.\,\ref{fig:EvoAfterInfla}).

\begin{figure}[!h!]
\begin{center}
$$\includegraphics[width=.47\textwidth]{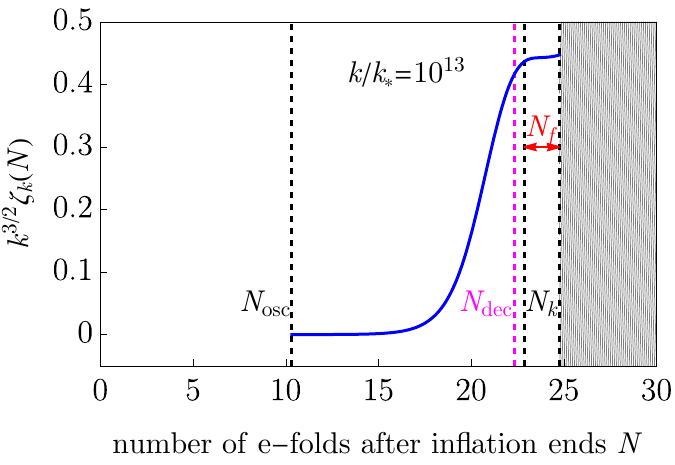}
\qquad\includegraphics[width=.47\textwidth]{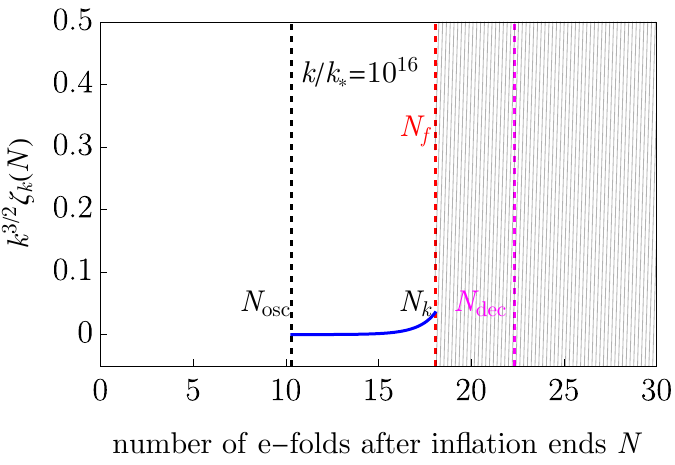}$$
\caption{\em   
Growth of the curvature perturbation for different modes. Left panel: the mode $k$ in exam re-enters the horizon at $N_k > N_{\rm dec}$ meaning that it still has some time to grow after the time of curvaton decay. This reflects in an enhancement in the amplitude of the power spectrum with respect to the sudden decay approximation. Right panel: the mode $k$ in exam re-enters the horizon before the curvaton decay and its value should be chosen at $N_f \simeq N_{k}$.
 }\label{fig:EvoZeta}  
\end{center}
\end{figure}

For each set of parameters, we show the power spectrum computed, as customary in the literature, in the sudden decay approximation (see ref.\,\cite{Ando:2017veq} for a detailed analysis). Additionally, in fig.\,\ref{fig:PS} we show that, by computing $P_\zeta(k)$ studying the full evolution of $\zeta_k$, one gets an enhancement of roughly a factor 2 in the amplitude with respect to the sudden decay approximation. 
This is a consequence of the fact that in the sudden decay approximation one computes the plateau of the power spectrum at $N_{\rm dec}$ but, as shown in fig.\,\ref{fig:EvoZeta}, $\zeta_k$ continues growing after that and then $\zeta_k(N_{\rm dec}) < \zeta_k(N_{\rm f})$.

It is worth noticing that, by properly tuning the initial parameters of the model, we are able to obtain a broad power spectrum which spans from the scales relevant to produce a signal compatible with NANOGrav (and other PTA experiments) to the ones at which PBHs can comprise the totality of dark matter, realising the scenario proposed in Ref.~\cite{DeLuca:2020agl} by means of an explicit particle physics model of inflation. 

Recently, loop corrections to the power spectrum of curvature perturbations in single-field models featuring an Ultra Slow-Roll (USR) phase have been intensively discussed in the literature~\cite{Kristiano:2022maq,
Cheng:2021lif,
Choudhury:2023hvf,
Choudhury:2023jlt,
Choudhury:2023rks,
Choudhury:2023vuj,
Riotto:2023gpm,
Riotto:2023hoz,
Firouzjahi:2023ahg,
Firouzjahi:2023aum,
Kristiano:2023scm,
Franciolini:2023lgy,
Tasinato:2023ukp}, after the suggestion such scenario would be incompatible with Cosmic Microwave Background (CMB) bounds on perturbations at large scales~\cite{Kristiano:2022maq}. 
In reference \cite{Franciolini:2023lgy}, it was however shown that, in realistic scenarios, carefully requiring $f_\text{\tiny PBH}\leq 1$ already force the USR modes 
to induce loop corrections which remain of the  order of a few percent, thus retaining the perturbativity of the theory. 
Since, in the curvaton scenario discussed here, the enhancement of curvature perturbations at small scales is due to the presence of a spectator field, while the power spectrum at the scales relevant for the CMB is governed by a decoupled sector, this scenario differs significantly the one discussed in the references above. Nonetheless, it would still be valuable to evaluate the potential impact of loop corrections within this framework. We leave this task for a future work.

\subsection{Primordial non-gaussianities}\label{sub:X}
Multi-field scenarios, as the curvaton model, can lead to sizeable NGs.
We follow ref.\,\cite{Sasaki:2006kq} and refs. therein. 
Requiring that the total energy density is uniform, conservation of energy allow us to relate the curvature perturbation to the energy density fraction of the curvaton
\begin{equation}
\label{eq:zeta_zetaG}
    e^{4 \zeta} - \Omega_\phi e^{3\zeta_\phi}e^\zeta-\left(1-\Omega_\phi\right)=0\,.
\end{equation}
Here we have also assumed, as customary in curvaton models, that the contribution to curvature perturbation coming from the radiation fluid is subdominant with respect to the curvaton one at small scales. By solving the above equation perturbatively, we are able to find an order-by-order relation between $\zeta$ and $\zeta_\phi$ where, in the flat gauge, 
\begin{equation}
    \zeta_\phi=\frac{1}{3}\log\left(1+\frac{\delta\rho_\phi}{\rho_\phi}\right)\,.
\end{equation}
After defining $\zeta_{\rm G} = r_{\rm dec} \zeta_{\phi}^{(1)}$, where $\zeta_{\phi}^{(1)}$ is the first-order term in the expansion, the power series for $\zeta_\phi$ can be resummed to the exact relation
\begin{equation}
    \zeta_\phi = \frac{2}{3}\log\left( 1+\frac{3}{2r_{\rm dec}}\zeta_{\rm G}\right)\,.
\end{equation}
By plugging the above relation into eq.\,(\ref{eq:zeta_zetaG}), and after defining 
\begin{align}\label{eq:MasterX}
    \zeta = \log\big[X(r_{\rm dec},\zeta_{\rm G})\big]\,,
\end{align}
we get the forth-order polynomial equation
\begin{equation}
    X^4 - \Omega_\phi\left(1+\frac{3}{2 r_{\rm dec}}\zeta_{\rm G}\right)^2 X - \left( 1-\Omega_\phi\right)=0\,,
\end{equation}
which can be solved to find
\begin{align}\label{eq:XFunction}
&X(r_{\rm dec},\zeta_{\rm G}) \equiv \frac{1}{\sqrt{2 (3+r_{\rm dec})^{1/3}}}
\Bigg\{
\sqrt{
\frac{
-3 + r_{\rm dec}(2+r_{\rm dec}) + [(3+r_{\rm dec})P(r_{\rm dec},\zeta_{\rm G})]^{2/3}
}{
(3+r_{\rm dec})P^{1/3}(r_{\rm dec},\zeta_{\rm G})
}
}+   
\nn \\
&
\sqrt{
\frac{(1-r_{\rm dec})}
{P^{1/3}(r_{\rm dec},\zeta_{\rm G})} -
\frac{P^{1/3}(r_{\rm dec},\zeta_{\rm G})}
{(3+r_{\rm dec})^{1/3}} 
+ 
\frac{
(2r_{\rm dec} + 3\zeta_G)^{2}
P^{1/6}(r_{\rm dec},\zeta_{\rm G})
}{r_{\rm dec}
\sqrt{-3 + r_{\rm dec}(2+r_{\rm dec}) +
[(3+r_{\rm dec})P(r_{\rm dec},\zeta_{\rm G})]^{2/3}
}}} 
\Bigg\}\,,
\end{align}
and 
\begin{align}
P(r_{\rm dec},\zeta_{\rm G}) \equiv 
\frac{(2r_{\rm dec} + 3\zeta_{\rm G})^{4}}{16 r_{\rm dec}^2} + \sqrt{
(1-r_{\rm dec})^3(3+r_{\rm dec}) + \frac{(2r_{\rm dec} + 3\zeta_{\rm G})^8}{256 r_{\rm dec}^4}
}\,.\label{eq:PFunction}
\end{align}
It will be useful in the following section to extract the leading order expansion of eq.\,(\ref{eq:MasterX}), 
which can be approximated at the second order in the Gaussian component $\zeta_{\rm G}$ as
\begin{align}\label{eq:ZetaQ}
\zeta_{2} \equiv \zeta_{\rm G} + \frac{3}{5}f_{\rm NL}(r_{\rm dec})\zeta_{\rm G}^2\,,
~~~~~~~{\rm with}~~~~~~~
f_{\rm NL}(r_{\rm dec}) \equiv  \frac{5}{3}\bigg(
\frac{3}{4r_{\rm dec}} - 1 - \frac{r_{\rm dec}}{2}
\bigg)\,.
\end{align}
It is typically assumed that computing the predictions of the model assuming only the leading order correction to the Gaussian case provides a sufficiently good approximation. 
However, we show in the next section that this is not the case when one calculates the abundance of PBHs, which are extremely sensitive to non-gaussian corrections. 

\section{Phenomenology:
\\primordial black holes and gravitational waves}\label{sec:Pheno}

In this section we investigate the phenomenological properties of the axion-curvaton scenario presented in this work.
In particular, we compute the PBH abundance as well as the SGWB associated to the enhanced perturbation at small scales, by fully accounting for the intrinsic primordial NGs imprinted in the curvature perturbations.
We show that by selecting specific values for the parameters of the model, it is possible to obtain a broad power spectrum that encompasses both the scales required for the totality of the dark matter in the form of PBHs and those relevant for a GW signal that is consistent with the one potentially hinted by PTA experiments. 
Moreover, we highlight that limiting our analysis to the quadratic approximation, while also accounting for primordial NGs, leads to incorrect phenomenological predictions when considering also the signal of GWs associated to PBHs formation.

\subsection{Primordial black hole
formation}

When computing the PBH abundance we will follow the general prescription presented in ref.\,\cite{PhysRevD.107.043520} (see also refs.\,\cite{Young:2022phe,Gow:2022jfb}), where more details can be found.
The approach is based on threshold statistics on the compaction function, a fundamental variable defined as twice the local mass excess over the areal radius 
\begin{align}\label{eq:DefinitionCompaction}
\mathcal{C}(r,t) &\coloneqq \frac{2\left[M(r,t) - M_b(r,t)\right]}{R(r,t)} 
\nn 
\\
&=
\frac{2}{R(r,t)}\int_{S^2_R} d^{3}\vec{x}
\left[\rho(\vec{x},t) - \rho_b(t)\right] = 
\frac{2}{R(r,t)}
\underbrace{\int_{S^2_R} d^{3}\vec{x}\,\rho_b(t)\,\delta(\vec{x},t)}_{\coloneqq\,\delta M(r,t)}\,.
\end{align}
Assuming spherical symmetry and adopting the gradient expansion approximation, on super-horizon scale the relation between the density contrast $\delta$ and the curvature perturbation $\zeta$ is non-linear\,\cite{Harada:2015yda,Musco:2018rwt} 
\begin{align}\label{eq:SphericalDelta}
\delta(r,t) = 
-\frac{2}{3}
\Phi
\left(\frac{1}{aH}\right)^2 
e^{-2\zeta(r)}\left[
\zeta^{\prime\prime}(r) + \frac{2}{r}\zeta^{\prime}(r) + \frac{1}{2}\zeta^{\prime}(r)^2
\right]\,.
\end{align}
The parameter $\Phi$ is introduced to keep track of how the equation of state of the Universe changes due to the thermail history, which is particularly relevant for the formation of stellar mass PBH across the QCD phase transition.
In case of a constant equation of state parameter $\omega = p/\rho$, its value takes the form
$\Phi = {3(1+\omega)}/{(5+3\omega)}$.
However, when $\omega$ retains a time dependence, deviations from the stationary solution are observed. In this work we adopt the fit derived in ref.~\cite{Franciolini:2022tfm}.

The non-linear nature of the above equation unavoidably introduces a certain amount of NGs in the PDF of $\delta$, and hence on that of $\mathcal{C}$. Using the above expression and integrating over the radial coordinate, 
eq.\,(\ref{eq:DefinitionCompaction}) takes the form
\begin{align}\label{eq:CompactionFull}
\mathcal{C}(r) = 
-2\Phi\,r\,\zeta^{\prime}(r)\left[
1 + \frac{r}{2}\zeta^{\prime}(r)
\right] = 
\mathcal{C}_1(r) - \frac{1}{4\Phi}\mathcal{C}_1(r)^2\,,
\end{align} 
where $\mathcal{C}_1(r)\coloneqq -2\Phi\,r\,\zeta^{\prime}(r)$ is defined as the linear component of the compaction function. The latter can be recast in terms of its Gaussian counterpart as follows
\begin{align}
    \mathcal{C}_1(r) = \mathcal{C}_{\rm G}(r)\,\frac{d F}{d\zeta_{\rm G}} \, ,~~~~~~~ \mathcal{C}_G(r) \coloneqq -2\Phi\,r\,\zeta^{\prime}_G(r)\,,
\end{align}
where $F$ encodes the relation between $\zeta$ and the Gaussian component $\zeta_{\rm G}$.
From eq.\,\eqref{eq:MasterX}, this is defined as $F= \log[X(r_{\textrm{dec}},\zeta_{G})]$.
We stress that $\mathcal{C}(r)$ depends both on $\zeta_{\rm G}$ and $\mathcal{C}_{\rm G}$ which are, by definition, gaussianly distributed.
Therefore, their joint PDF can be computed as \begin{equation}
\mathrm{P}_{\mathrm{G}}\left(\mathcal{C}_{\rm G}, \zeta_{\mathrm{G}}\right)=\frac{1}{2\pi\sigma_{c} \sigma_{r} \sqrt{1-\gamma_{c r}^{2}}} \exp \left(-\frac{\zeta_{\mathrm{G}}^{2}}{2 \sigma_{r}^{2}}\right) \exp \left[-\frac{1}{2\left(1-\gamma_{c r}^{2}\right)}\left(\frac{\mathcal{C}_{\mathrm{G}}}{\sigma_{c}}-\frac{\gamma_{c r} \zeta_{\mathrm{G}}}{\sigma_{r}}\right)^{2}\right],
\end{equation} 
where $\gamma_{cr}\equiv\sigma_{cr}^2/\sigma_c \sigma_r$ and the correlators are given by
 \begin{align}
 \langle\mathcal{C}_{\rm G}\mathcal{C}_{\rm G}\rangle & = \sigma_c^2 =
  \frac{4\Phi^2}{9}\int_0^{\infty}\frac{dk}{k}
  (kr_m)^4 W^2(k,r_m) T^2(k,r_m) P_{\zeta}(k)\,,\label{eq:Var1}
   \\
 \langle\mathcal{C}_{\rm G}\zeta_{\rm G}\rangle & = \sigma_{cr}^2 = 
 \frac{2\Phi}{3}\int_0^{\infty}\frac{dk}{k}(kr_m)^2
 W(k,r_m)
 W_s(k,r_m) T^2(k,r_m) P_{\zeta}(k)\,,
  \\
  \langle\zeta_{\rm G}\zeta_{\rm G}\rangle & = \sigma_r^2 =   \int_0^{\infty}\frac{dk}{k}
  W_s^2(k,r_m) T^2(k,r_m) P_{\zeta}(k)\,,\label{eq:Var3}
 \end{align}
 with $W_s(k,r) = \sin(kr)/kr$ while $W(k,R)$ and $T(k,\tau)$ are given by
 \begin{align}
W(k,R) &= 3\big[
\frac{\sin(kR) - kR\cos(kR)}{(kR)^3}\big]\,,
\quad
\nn \\
T(k,\tau) &= 3\big[
\frac{\sin(k\tau/\sqrt{3}) - (k\tau/\sqrt{3})\cos(k\tau/\sqrt{3})}{(k\tau/\sqrt{3})^3}\big]\,.  \label{eq:T}
 \end{align}
The transfer function $T(k,\tau)$ we adopt in this section is derived assuming a perfect radiation fluid. Near the QCD epoch, the softening of the equation of state slightly affects the evolution of modes within the sub-horizon regime. 
We will not include this effect here, as it is mitigated by the adoption of a window function that also smooth out sub-horizon modes, while it only concerns a tail of the PBH mass function around the solar mass. We will discuss in the next section how this effect may be more important on the shape of the SGWB spectrum within the PTA frequency range. 
The computation of PBHs NG abundance from the collapse of a single mode is finally
\begin{align}
\beta_{\rm NG} & = \int_{\mathcal{D}}\mathcal{K}(\mathcal{C} - \mathcal{C}_{\rm th})^{\gamma}
\textrm{P}_{\rm G}(\mathcal{C}_{\rm G},\zeta_{\rm G})d\mathcal{C}_{\rm G} d\zeta_{\rm G}\,,\label{eq:CompactionIntegral}
\end{align}
with
\begin{align}
 \mathcal{D} & = 
\left\{
\mathcal{C}_{\rm G},\,\zeta_{\rm G} \in \mathbb{R}~:~~
\mathcal{C}(\mathcal{C}_{\rm G},\zeta_{\rm G}) > \mathcal{C}_{\rm th}  
~\land~\mathcal{C}_1(\mathcal{C}_{\rm G},\zeta_{\rm G}) < 2\Phi
\right\}\,.\label{eq:RegionD}
\end{align}

The final mass distribution of PBHs at the end of formation era can be obtained by integrating over all epochs when the formation is active, corresponding to epochs of horizon crossing of momenta $k$ where the curvature power spectrum is enhanced. 
For convenience, we express the horizon mass $M_H$ in terms of the related power spectral modes through\,\cite{Franciolini:2022tfm}
\begin{equation}
M_H \simeq 17 M_{\odot}\left(\frac{g_*}{10.75}\right)^{-1 / 6}\left(\frac{k / \kappa}{\mathrm{pc}^{-1}}\right)^{-2}
\end{equation}
where the parameter $\kappa$ relates the size of the overdensity to the $r_m$ (that crosses the horizon when its mass is $M_H$) to the momentum $k$.
We will fix this parameters below, adopting the results of numerical simulations \cite{Musco:2020jjb}.
The PBH abundance $f_{\rm PBH}$ can be 
derived from the mass fraction $\beta_{\rm NG}$ as
\begin{align}
f_{\rm PBH}(M_{\rm PBH}) 
\equiv
\frac{1}{\Omega_{\rm  DM}} 
\frac{d\Omega_{\rm PBH}}
{d\log M_{\rm PBH}}\,,
~~~~
{\rm with}
~~~~
\Omega_{\rm PBH} = 
\int
d \log M_{H} \left(\frac{M_{\rm eq}}{M_{H}} \right)^{1/2}\beta_{\rm NG}(M_{H})\,,
\label{eq:diffmassfraction}
\end{align}
where $M_{\rm eq} \approx 2.8\times 10^{17}\,\,M_{\odot}$ is the horizon mass at the time of matter-radiation equality and $\Omega_{\rm  DM}$ is the cold dark matter density of the universe ($\Omega_{\rm  DM} \simeq 0.12\,h^{-2}$ with $h = 0.674$ for the Hubble parameter).
Imposing that only over-threshold perturbations ${\cal C} \geq {\cal C}_{\rm th}$ eventually collapse to form PBHs, and using eq.\,\eqref{eq:CompactionFull}, one finds that the critical values of the linear component ${\cal C}_\text{G}$ corresponds to 
\begin{align}
    {\cal C}_{\text{G}, {\rm th}, \pm} = 
    2\Phi \left(\frac{dF}{d\zeta_{\rm G}} \right)^{-1}
    \left(1 \pm \sqrt{1-\frac{{\cal C}_\text{th}}{\Phi}}\right)\,,
\end{align}
where we choose the minus sign  as we focus only on type-I perturbations\,\cite{Musco:2020jjb}. Thus, the integration range is
\begin{align}
     {\cal C}_{\text{G}, {\rm th}, - } \leq  {\cal C}_{\text{G}} \leq 2\Phi \left(\frac{dF}{d\zeta_{\rm G}} \right)^{-1}.
\end{align}
The mass of the PBH formed follows the properties of critical collapse, and can be related to the horizon mass $M_H$ using
\begin{equation}
M_{\rm PBH} = \mathcal{K}M_H\left[
    \left({\cal C} - \frac{1}{4\Phi}{\cal C}^2\right)
    - {\cal C}_{\rm th}
    \right]^{\gamma},
\end{equation}
where $\mathcal{K}$ and $\gamma$ are parameters obtained by numerical simulations of the collapse.
The above expression can be inverted as 
\begin{equation}
{\cal C}_\text{G}  = 
2\Phi 
\left(\frac{dF}{d\zeta_{\rm G}} \right)^{-1} 
\left[
1 - \sqrt{1 - \frac{{\cal C}_{\rm th}}{\Phi} - 
\frac{1}{\Phi}\left(
 \frac{M_{\rm PBH}}{\mathcal{K}M_H}
\right)^{1/\gamma}}
    \right]\,,
\end{equation}
so that now we can express the integration over $ d{\cal C}_{\rm G} d \zeta_{\rm G}$ in eq.\,\eqref{eq:CompactionIntegral} in terms of $d{M}_{\rm PBH}d \zeta_{\rm G}$, and then consider the differential mass fraction \eqref{eq:diffmassfraction}. 
Finally, the overall late-time universe abundance of PBH of given mass $M_{\rm PBH}$ comes out of the integration over the allowed horizon masses $M_H$, i.e. over the possible epochs of formation. We obtain
\begin{align}
 f_{\rm PBH}(M_{\rm PBH})
 & =
 \frac{1}{\Omega_{\rm DM}}
 \int_{M_{H}^{\rm min}(M_\text{PBH})}
 d \log M_{H} \left(\frac{M_{\rm eq}}{M_{H}}\right)^{1/2}
 \left[
 1 - \frac{{\cal C}_{\rm th}}{\Phi} - 
\frac{1}{\Phi}\left(
 \frac{M_{\rm PBH}}{\mathcal{K}M_H}
\right)^{1/\gamma} 
\right]^{-1/2}
\nn \\
& \times 
  \frac{{\cal K}}{\gamma}
 \left(\frac{M_{\rm PBH}}{\mathcal{K} M_{H}}\right)^\frac{1+\gamma}{\gamma}
 \int d\zeta_\text{G}
 P_{\rm G}({\cal C}_{\rm G}(M_{\rm PBH},\zeta_\text{G}), \zeta_\text{G}|M_{H})
\left(\frac{dF}{d\zeta_{\rm G}} \right)^{-1}  ,
\end{align}
\noindent
where the integrand also includes the determinant of the Jacobian and the horizon mass dependence of 
the PDF $P_{\rm G}(M_{H})$ 
is inherited by the smoothing scale $r_m(M_H)$ controlling the variances 
$(\sigma_{c},\sigma_{cr}, \sigma_{r})$, fixing the horizon crossing epoch.
In this work, we have followed the prescription given in ref.\,\cite{Musco:2020jjb} to compute the values of $\mathcal{C}_{\rm th}$ and $r_m$\footnote{Notice that the prescription in ref.\,\cite{Musco:2020jjb} to compute the threshold for PBH collapse only accounts for NGs arising from the non-linear relation between the curvature perturbations and the density contrast.  
In principle, also primordial NGs beyond the quadratic approximation should be taken into account when computing the threshold value (see e.g. refs.~\cite{Kehagias:2019eil,Escriva:2022pnz} for works in this direction). This is left for future work.}, which depend on the shape of the power spectrum.
As the power spectrum we obtained in the axion-curvaton scenario we consider in this work is nearly scale-invariant, one gets $\mathcal{C}_{\rm th}=0.55$ and $kr_m= 4.49\equiv \kappa$.
The presence of the QCD phase transitions is taken into account by considering, as shown in refs.\,\cite{Franciolini:2022tfm,Muscoinprep}, that $\gamma(M_H)$, $\mathcal{K}(M_H)$, $\mathcal{C}_{th}(M_H)$ and $\Phi(M_H)$ are functions of the horizon mass for $M_{\rm PBH}=\mathcal{O}(M_{\odot})$.

\begin{figure}[!h!]
\begin{center}
$$\includegraphics[width=.40\textwidth]{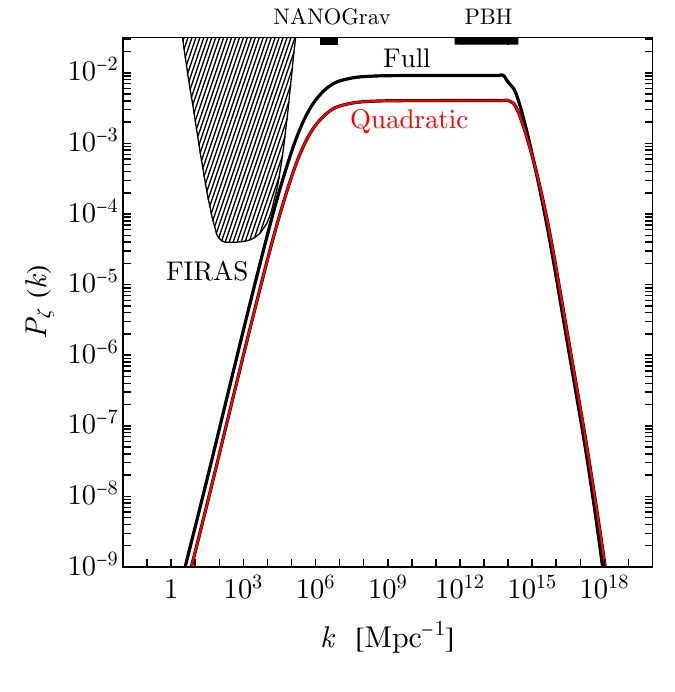}
~~\includegraphics[width=.59\textwidth]{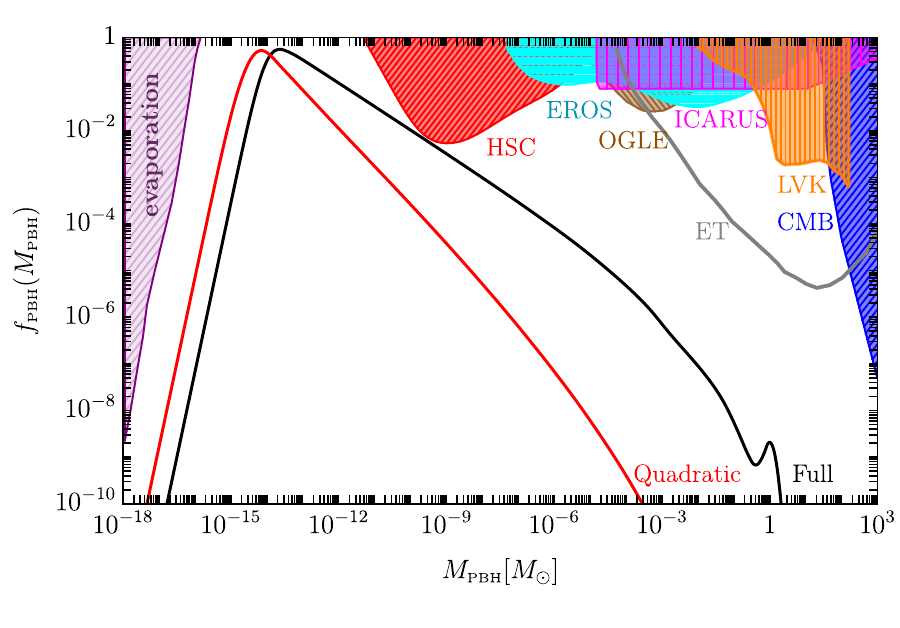}$$
\caption{\em
Left panel. Broad power spectrum of the curvature perturbation obtained with the axion-curvaton model assuming $N_{*}=58$. The black solid line refers to the power spectrum needed to obtain $f_{\rm PBH}\simeq 1$ when computing the abundance of PBHs with a non-perturbative treatment of NGs. 
We chose $m_{\phi}=5.15\times10^7$ ${\rm GeV}$,\;$f=4.20\times10^{13}$ ${\rm GeV}$,\;$\vartheta_0=5.00\times 10^{-2}$. 
For comparison, we also show an analogous scenario (red line) providing $f_{\rm PBH}=1$  in the quadratic approximation and corresponding to the choice of $m_{\phi}=1.00\times10^8$ ${\rm GeV}$$,\;f=6.30\times10^{13}$ $ {\rm GeV,\;\vartheta_0=5.00\times 10^{-2}}$. For both cases we get $r_{\rm dec}=0.5$. The shaded region shows the experimental
constraints coming from the analysis of CMB spectral distortion by the FIRAS collaboration\,\cite{Fixsen:1996nj,Bianchini:2022dqh}. Right panel. $f_{\rm PBH}(M_{\rm PBH})$ computed starting from the corresponding power spectrum in the Full (black solid line) and in the quadratic (red dashed line) computation, as in the left panel. In both cases PBHs comprise all the dark matter in the Universe. 
Constraints on $f_{\rm PBH}$ shown in the plot are addressed in the main body of section\,\ref{sec:Pheno}.
 }\label{fig:TypeIDM}  
\end{center}
\end{figure}

The integrated abundance of PBHs is given by the integral 
\begin{equation}
f_{\rm PBH}=\int f_{\rm PBH}(M_{\rm PBH})d\log M_{\rm PBH}.
\end{equation}
We tune the parameters of the axion-curvaton model 
requiring PBHs to account for the totality of dark matter, i.e. fixing $f_{\rm PBH}\simeq1$.
It is instructive to compute the mass fraction of PBHs including NGs both with the quadratic approximation (eq.\,\eqref{eq:ZetaQ}) and the exact functional form (eq.\,(\ref{eq:MasterX})).
This allows us to investigate the relevance of the non-perturbative treatment of NGs which we adopt here based on ref.~\cite{PhysRevD.107.043520}.
Fig.\,\ref{fig:TypeIDM} shows $f_{\rm PBH}$ so computed, together with the corresponding power spectrum obtained through the mechanism presented in section\,\ref{sec:TypeI}.

In fig.~\ref{fig:TypeIDM} one can see that the power spectrum of curvature perturbations grows from its small value at large scales to the enhanced plateau at $k\gtrsim 10^6$ Mpc$^{-1}$. We choose $c = 1.6$ in eq.~\eqref{eq:EffectiveV}, which implies a
power law growth ${\cal P}_\zeta\approx k^{n_\theta}$ 
of the curvature spectrum
with an index  $n_\theta = 1.4$. 
It is interesting to notice that this growth is shallower than what it is typically achieved in single field models of inflation characterised by an USR phase\,\cite{Byrnes:2018txb}.

In the right panel of fig.\,\ref{fig:TypeIDM} we show the most stringent experimental constraints on $f_{\rm PBH}$ (see ref.\,\cite{Green:2020jor} for a review and\,\href{github.com/bradkav/PBHbounds}{\faGithub/bradkav/PBHbounds}). Constraints from evaporation include (see also\,\cite{Saha:2021pqf,Laha:2019ssq,Ray:2021mxu}): CMB\,\cite{Clark:2016nst}, EDGES\,\cite{Mittal:2021egv},  INTEGRAL\,\cite{Laha:2020ivk,Berteaud:2022tws}, Voyager\,\cite{Boudaud:2018hqb}, 511\;keV\,\cite{DeRocco:2019fjq},
EGRB\,\cite{Carr:2009jm}; HSC (Hyper-Supreme Cam)\,\cite{Niikura:2017zjd}, EROS\,\cite{EROS-2:2006ryy}, OGLE\,\cite{Niikura:2019kqi} and Icarus\,\cite{Oguri:2017ock} constraints come from microlensing-related observations\footnote{
Note that the development of dark matter halos dressing PBHs may strengthen the microlensing constraints\,\cite{Cai:2022kbp}. 
However, this effect is sensitive to assumptions on the relation between the mass of the PBH and the one of the surrounding halo, as well as on the assumed radius of the halo itself. We leave the investigation of this effect in scenarios where asteroidal mass PBH compose the entirety of the dark matter for future work.}; constraints coming from modification of the CMB spectrum due to accreting PBHs are derived in ref.\,\cite{Serpico:2020ehh} (see also ref.\,\cite{Piga:2022ysp}); the range around $M_{\odot}$ is constrained by LIGO observations on PBH-PBH merger, as recently derived in ref.\,\cite{Franciolini:2022tfm} (see also\,\cite{Kavanagh:2018ggo,Hall:2020daa,Wong:2020yig,Hutsi:2020sol,DeLuca:2021wjr,Franciolini:2021tla}), while in gray we indicate the future bound that next generation of ground based detectors will set on the PBH abundance \cite{DeLuca:2021hde,Pujolas:2021yaw,Franciolini:2022htd,Martinelli:2022elq,Franciolini:2023opt,Branchesi:2023mws}.

In accordance with what was observed in ref.\,\cite{PhysRevD.107.043520}, 
we find that for $r_{\rm dec}=0.5$, 
the effect of primordial NGs truncated to second order is that of enhancing PBHs abundance with respect to the exact computation. Therefore, requiring no PBH overproduction (that would surpass the dark matter abundance) imposes to decrease the amplitude of the power spectrum in this approximate case. 
Such a difference between the approximated and exact spectra
has an important impact on the signal of second-order GWs and could in principle be explored at GWs detectors, as shown in the following section.

It is important to note that, as we can see from the right plot of fig.\,\ref{fig:TypeIDM}, there is a second peak in $f_{\rm PBH}(M_{\rm PBH})$ around solar masses, 
which is caused by the softening of the equation of state during the QCD cross-over phase transition\,\cite{Jedamzik:1998hc,Byrnes:2018clq,Franciolini:2022tfm,Escriva:2022bwe,Muscoinprep}.
While this enhancement may be sizeable, it is not sufficient to generate a large enough abundance leading to a sizeable merger rate of stellar mass PBH mergers in this scenario. 
In practice, the properties of this model, in combination with the FIRAS bound that force the enhancement of perturbations to be placed at scales larger than around $\approx 10^5 / {\rm Mpc}$, cause the PBH abundance in the stellar mass range to be small.\footnote{Changing the assumed value of $c$ to larger numbers, i.e. larger $n_\theta$, would slightly alleviate such a conclusion due to a steeper spectral growth around $10^4$/Mpc.} 
Furthermore, the reason why the QCD peak is only visible in the exact computation has to be found in the left-hand plot in fig.\,\ref{fig:TypeIDM}. The quadratic approximation leads to an overproduction of PBHs, which has to be compensated by a decrease in the amplitude of the power spectrum if one wants not to overshoot the limiting value $f_{\rm PBH}=1$. 
Readjusting the parameters to decrease the amplitude of the power spectrum leads to a slight shift of the rising slope towards larger $k$, resulting in a smaller abundances at high masses where the QCD transition would have an impact. 
 
\subsection{Scalar-induced gravitational waves}\label{Sec:SGWB}

PBHs formation occurs as large curvature perturbations re-enter the Hubble horizon after inflation and eventually collapse under the effect of gravity. The same enhanced scalar perturbations emit tensor modes 
thanks to second-order effects around the epoch of horizon crossing. This generates an observable SGWB (see ref.\,\cite{Domenech:2021ztg} for a recent review). 

We compute the emission of GWs by accounting for the softening of equation of state at the QCD era, which has an important role in shaping the spectral tilt in the PTA frequency range, following ref.\,\cite{Abe:2020sqb}.
On the other hand, we will neglect higher order contributions to the SGWB. This is because, differently from what happens in the case of PBH formation which are extremely sensitive to the non-gaussian tail of the curvature distribution, 
the emission of GWs is dominated by the leading order in our case. Indeed, in the scenario we consider, we have $r_{\rm dec} = 0.5$ that corresponds to $f_{\rm NL}= 0.42$ from eq.\,\eqref{eq:ZetaQ}. 
Therefore, as shown in refs.\,\cite{Cai:2018dig,Unal:2018yaa,Ragavendra:2021qdu,Adshead:2021hnm,Abe:2022xur,Garcia-Saenz:2022tzu}, 
corrections from higher orders terms only amount to a negligible contribution to the SGWB.

The emission of GWs is dictated by the second order equation\,\cite{Tomita:1975kj,Matarrese:1993zf,Acquaviva:2002ud,Mollerach:2003nq,Ananda:2006af,Baumann:2007zm}
\begin{equation}
\left [ \frac{d^2}{d \eta^2} + k^2 - \left ( \frac{1-3w(\eta)}{2}  \right) 
{\cal H}^2 \right ]
 a(\eta)h _{\vec k}(\eta) 
= 4 a(\eta) {\cal S}_{\vec k}(\eta),
    \label{eq: EoM_gw}
\end{equation}
where $\eta$ is the conformal time, ${\cal H}\equiv a H$ is the conformal Hubble parameter, and the source term is written in terms of the gravitational potential $\Phi$ the conformal Newtonian gauge as 
\begin{align}
{\cal S}_{\vec k} =& \int \frac{\text{d}^3 q}{(2 \pi)^{3}} 
e_{ij}({\vec k }) q_i q_j 
\left[ 2\Phi_{\vec q}  \Phi_{\vec k -\vec q} 
+ \frac{4}{3(1+w)} 
\left( \mathcal{H}^{-1} \Phi'_{\vec q} + \Phi_{\vec q}\right) \left( \mathcal{H}^{-1} \Phi'_{\vec k -\vec q}  + \Phi_{\vec k -\vec q}  \right)  \right].
\end{align}
The evolution of the scalar perturbations is modified by the softening of the equation of state around the QCD era. For this reason, we solve numerically the evolution of $\Phi_{\vec k}$ given by (see e.g. ref.\,\cite{Mukhanov:2005sc})
\begin{align}
\Phi''_{\vec k} + 3 \mathcal{H} (1 + c_{\text{s}}^2) \Phi'_{\vec k} 
+ [
2 \mathcal{H}'+(1+3 c_{\text{s}}^2)\mathcal{H}^2 +c_{\text{s}}^2 k^2
] \Phi_{\vec k}  = 0, 
\label{EOM_Phi_complete}
\end{align}
only for spectral modes re-entering the horizon close to the QCD phase transition around $k \simeq 10^6$ Mpc$^{-1}$, 
while using the analytical solution assuming perfect radiation (i.e. eq.\,\eqref{eq:T}) otherwise.
Adopting the Green's function method, we solve for the tensor modes $h_{\vec k}$, accounting for the time-varying equation of state. 
The power spectrum of tensor modes becomes
\begin{align}
P_h (\eta, k) =  2 
 \int_0^\infty \text{d}t \int_{-1}^{1}\text{d} s 
 \left [ \frac{t(2+t)(s^2-1)}{(1-s+t)(1+s+t)} \right ]^2
 {\cal I}^2(t,s,\eta,k) 
 \nonumber \\
 \times P_\zeta \left ( k (t+s+1)/2 \right )
 P_\zeta \left ( k (t-s+1)/2 \right ),
 \label{P_h_ts}
\end{align}
and scales like the two powers of $P_\zeta$, due to the second order nature of the emission. 
We denote $P$ as the dimensionless power spectrum, following the convention adopted in the rest of this work. The kernel function ${\cal I}(t,s,\eta,k)$ is computed by integrating over time the Green's function multiplied by the time-dependent source (see more details in ref.\,\cite{Abe:2020sqb}).

The current abundance of SGWB can be computed accounting for the propagation as free GWs after emission, whose energy density in the late time universe is sensitive to the 
deviations from exact radiation dominated background due to the time dependence of $g_*$ and $g_{*s}$. One finds\,\cite{Espinosa:2018eve,Kohri:2018awv}
\begin{equation}
    \begin{aligned}\label{eq:OmegaGW}
        \Omega_{\rm GW}(k)h^2 
        =\Omega_{r,0}h^2
        \left (\frac{a_{\rm c}{\cal H}_{\rm c}}{a_{\rm f}{\cal H}_{\rm f}} \right )^2
        \frac{1}{24}\left (\frac{k}{{\cal H}_{\rm c}}\right )^2
        \overline{ P_h(k,\eta_{\rm c})}.
    \end{aligned}
\end{equation}
where
$\Omega_{r,0}$ stands for the current radiation density if the neutrino were massless and
we denoted as $\eta_c \gg 1/k $ the time after which GW emission of a given mode $k$ becomes negligible. For concreteness, following the choice of ref.\,\cite{Abe:2020sqb}, we fix $\eta_c =400/k $.

Two different effects modulates the SGWB around the nano-Hz frequencies, beyond what is expected from our ${ P}_\zeta$ in a pure radiation background.
The pre-factor $\left ({a_{\rm c}{\cal H}_{\rm c}}/{a_{\rm f}{\cal H}_{\rm f}} \right )^2 = \left ({g_*}/{g_*^0} \right) \left({g_{*S}^0}/{g_{*S}}\right)^{4/3} $ 
is typically denoted  $c_g$ in the literature.
This accounts for the departure of the cosmological expansion from the solution in perfect radiation when there is a variation of effective relativistic degrees of freedom. In practice, it tracks the different  dilution of the energy density in the GW sector compared to the background, which is particularly relevant across the QCD phase. 
On top of this, the smaller $c_s$ encountered around the QCD era delays the oscillation of density perturbations right after its horizon re-entry, compared to pure radiation dominated background, as a consequence of the smaller sound horizon $c_s/H$. 
This leads to an enhancement of the SGWB around $f\approx {\rm few}$ nHz, a feature right in the frequency range observed by PTA experiments.

\begin{figure}[!t!]
	\begin{center}
$$\includegraphics[width=.49\textwidth]{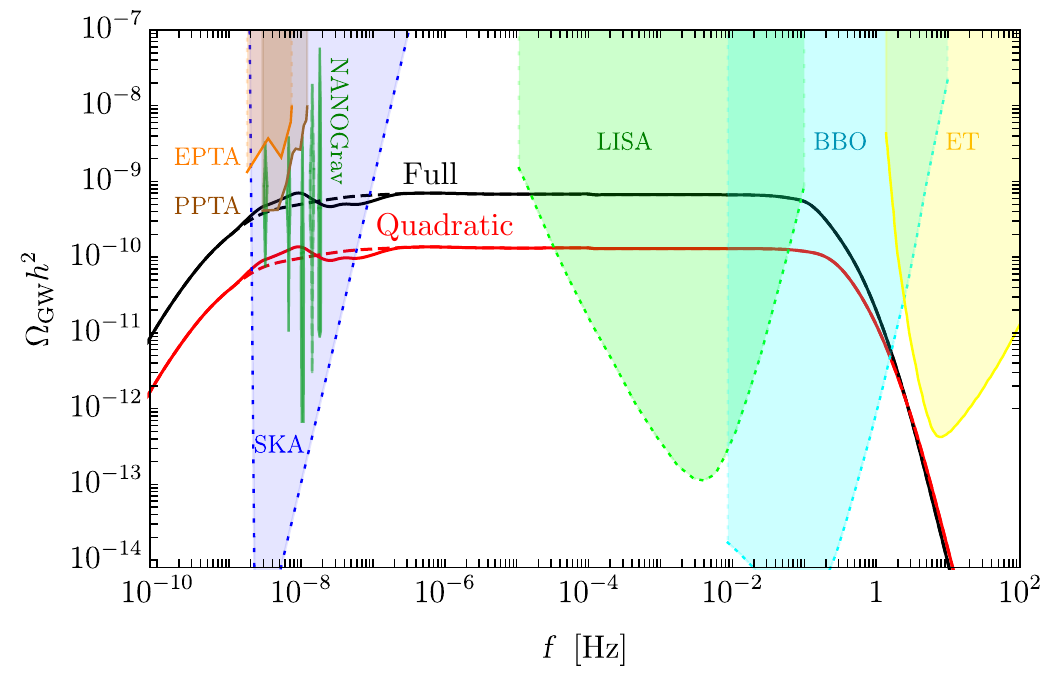}
~~\includegraphics[width=.49\textwidth]
{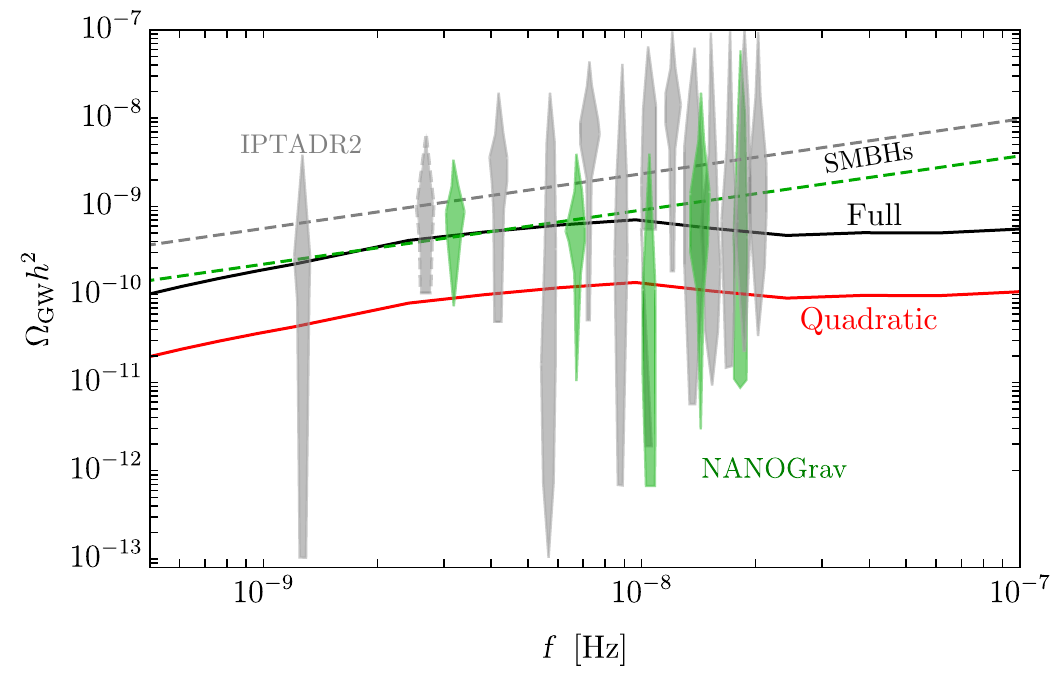}$$
		\caption{\em   
		Signal of second order GWs associated with the broad power spectrum obtained within the curvaton model. 
		When computing abundance of PBHs in the quadratic approximation (red line) and requiring $f_{\rm PBH}\simeq 1$, 
		one would find an amplitude which is smaller than the one required to explain the tentative signal detected by the NANOGrav and IPTA collaborations. 
		However, correctly adopting the non-perturbative computation of the PBH abundance from ref.~\cite{PhysRevD.107.043520} (black line), we find that it is possible to explain at the same time dark matter in the form of asteroidal mass PBHs and the SGWB signal from NANOGrav and IPTA within the curvaton scenario. 
		We plot constraints on the signal of second-order GWs coming from NANOGrav 12.5 yrs experiment\,\cite{NANOGrav:2020bcs},
  IPTA-DR2 \,\cite{Antoniadis:2022pcn}, EPTA\,\cite{Lentati:2015qwp}, PPTA\,\cite{Shannon:2015ect}, and future sensitivity for planned experiments like SKA\,\cite{Zhao:2013bba}, LISA\,\cite{LISA:2022kgy}, BBO/DECIGO\,\cite{Yagi:2011wg} and ET (power law integrated sensitivity curves as derived in ref.~\cite{Bavera:2021wmw}). 
		The dashed line report the SGWB obtained neglecting the variation of sound speed during the QCD era (around $f\approx 10^{-8}$Hz) but accounting for a temperature dependent overall factor $c_g$.
}\label{fig:GW}  
	\end{center}
\end{figure}

Finally, we compute the signal of GWs associated with the abundance of PBHs in fig.\,\ref{fig:TypeIDM}, by plugging the corresponding power spectrum into eq.\,(\ref{eq:OmegaGW}). Results are shown in fig.\,\ref{fig:GW}, where $\Omega_{\rm GW}h^2$ is given as a function of the frequency $f= k/2\pi$.\footnote{Notice that before the curvaton decays, its isocurvature perturbations may also induce second order GWs\,\cite{Bartolo:2007vp,Kawasaki:2013xsa}. We neglect the isocurvature contribution and only focus on the adiabatic source, as it would only affect the tail at large frequencies of the SGWB, while also being suppressed by powers of the small ratio $\Gamma_\phi/m_\phi$ considered in this work (see eq.~\eqref{eq:Gamma}).}
This plot shows that the curvaton scenario discussed in this work is able to produce an enhanced and flat power spectrum, that interestingly connects the asteroidal mass PBH dark matter and SGWB at PTA frequencies, providing a concrete realisation
of the scenario proposed in ref.~\cite{DeLuca:2020agl}.

One of the take-home messages of this paper is also contained in  fig.\,\ref{fig:GW}, that shows how primordial NGs have important phenomenological relevance and must be considered with care. 
Indeed, relying on the quadratic approximation when computing the PBH abundance would already exclude the possibility of explaining both the totality of dark matter and the tentative signal by NANOGrav 12.5 years observations, and other PTA datasets, within a unified PBH formation scenario based on the curvaton model.
On the other hand, as we have shown, the non-perturbative treatment of the NGs inherently induced in curvaton scenarios invalidates this conclusion. 
As a by-product of this result, we also show that 
the GW amplitude of PBH dark matter in the LISA frequency band is also modified when considering the specific NG formation scenario discussed in this work.

Recently, Ref.~\,\cite{Dandoy:2023jot} (see also ref.~\cite{Inomata:2018epa,Cai:2019elf,Chen:2019xse,Vaskonen:2020lbd, DeLuca:2020agl,Zhao:2022kvz,Yi:2022ymw}) conducted a Bayesian analysis of the latest PTA datasets to investigate whether a SGWB generated by curvature perturbations could be compatible with the recently reported NANOGrav 12.5 \cite{NANOGrav:2020bcs} and IPTA-DR2 \cite{Antoniadis:2022pcn} excess, assuming it is generated by GWs.
Furthermore, in their study, they compared the induced GW scenario with the leading astrophysical source candidate, i.e. supermassive black hole binaries (SMBHBs), and found that a cosmological interpretation of the SGWB provides a competitive explanation of the NANOGrav signal, while currently being disfavored by IPTA observations over the latter.\footnote{
Other cosmological candidates to explain the putative PTA signal are: annihilation of cosmic domain walls~\cite{Vilenkin:1981zs,Ferreira:2022zzo},
first order phase transitions~\cite{Witten:1984rs,NANOGrav:2021flc,Xue:2021gyq,Nakai:2020oit,Ashoorioon:2022raz,Benetti:2021uea,Hindmarsh:2022awe}, 
cosmic strings \cite{Blasi:2020mfx,Ellis:2020ena,Blanco-Pillado:2021ygr}, magnetic fields \cite{Neronov:2020qrl,RoperPol:2022iel}
and others \cite{Bian:2020urb,Vagnozzi:2020gtf}.
}
Overall, barring existing systematics in the computation of the abundance, it is found that PTA upper bounds are currently compatible with the condition $f_\text{PBH}\leq 1$.
We notice, however, that their analysis is performed assuming Gaussian curvature perturbations and peaked power spectra parameterised by a lognormal shape with a variable width.
As we discuss in this work, including primordial NGs (here derived within the curvaton model) in the calculation of the abundance of PBHs can significantly affect the amplitude of the power spectrum in order to explain the totality of dark matter, modifying the relation with upper constraints presented in\,\cite{Dandoy:2023jot}.
In the example presented in this section, with $r_{\rm dec} = 0.5$, the spectral amplitude required to obtain $f_{\rm PBH} = 1$ is close to the one obtained with Gaussian perturbations (and the only including the effect on non-linearities), while it would be reduced if smaller values of $r_{\rm dec}$ were considered (see also \cite{PhysRevD.107.043520}).

Finally, it is interesting to notice that, as can be observed from the right panel of fig.\,\ref{fig:GW}, the model examined in this work can produce a SGWB signal at nHz frequencies that deviates from the typical signal generated by the merger of SMBHBs, represented by dashed lines, assuming quasi-circular orbits driven by GW emission\,\cite{Phinney:2001di}.\footnote{We note, however, that several studies 
(e.g.\,\cite{Haiman:2009te,Kocsis:2010xa, Burke-Spolaor:2018bvk, Middleton:2020asl,Pan:2021oob,Ellis:2023owy}) based on a SMBHBs population synthesis codes predict a signal with a 
different slope compared to the result depicted by the dashed lines $\Omega_{\rm GW}\sim f^{2/3}$. }
In particular, beyond the small modulation induced by the QCD cross-over effect, the plateau is reached at frequencies larger than around $20$ nHz, where current PTA data are still dominated by pulsar intrinsic noise (and not shown for clarity).
We expect a clearer understanding of whether the supposed SGWB signature may be due to astrophysical or primordial sources will be reached with extended datasets and improved intrinsic noise modelling by PTA collaborations (see e.g.~\,\cite{NANOGrav:2020spf} for a roadmap).

\section{Conclusions}\label{sec:Finale}

The recent detection of GWs has opened up a new window to investigate the composition and properties of our universe.
The goal of this paper was two-fold. 
Firstly, we aimed to demonstrate that it is feasible to generate a broad enhanced power spectrum of the curvature perturbations in a specific curvaton model. This was achieved by carefully selecting the initial parameters of the model, which resulted in a broad power spectrum that encompasses both the scales required for the totality of the dark matter in the form of PBH and those relevant for GW signal that is consistent with the one potentially hinted by NANOGrav and other PTA experiments.
Secondly, we showed in a concrete example the phenomenological relevance of going beyond perturbative approaches when computing the PBH abundance in presence of NGs.

We utilized linear cosmological perturbation theory to calculate the dynamics of curvature perturbations. 
By carefully analyzing the evolution of each mode, $\zeta_k$, we were able to compute the power spectrum, $P_\zeta(k)$, going beyond the sudden decay approximation. We found that $\zeta_k$  continues to grow after the curvaton decay, resulting in an enhancement of approximately a factor of $\mathcal{O}(2)$ in the amplitude of $P_\zeta(k)$ compared to the sudden decay approximation, which has important phenomenological consequences.

When computing the abundance of PBHs, we utilized the technique presented in ref.~\,\cite{PhysRevD.107.043520}, which accounts for both NGs induced by non-linearities and primordial NGs, predicted in the curvaton model, and encoded in the exact expression in eq.~\,(\ref{eq:MasterX}). We compared this result to approximations typically employed in the literature, which only include primordial NGs based on perturbative expansion truncated at the quadratic order. 
We found that the quadratic approximation of NGs leads to  an overestimation of the PBH abundance compared to the results obtained using the full non-Gaussian relation, 
due to the relevant impact of higher-order terms and a violation of perturbativity in the computation of PBH abundance observed for broad spectra \cite{PhysRevD.107.043520}. 
This leads to an interesting phenomenological outcome: the amplitude of $\Omega_{\rm GW}$ obtained by considering the exact relation $\zeta\left(\zeta_{\rm G}\right) = \log\left[X\left(r_{\rm dec},\zeta_{\rm G}\right)\right]$ can better fit the tentative signal observed by NANOGrav and IPTA, as demonstrated in fig.\,\ref{fig:GW}, only if the NGs are correctly computed beyond the quadratic approximation, highlighting the importance of a precise calculation of the impact of non-Gaussian corrections when considering both PBH abundance and scalar-induced GWs.
We expect the next data release by PTA collaborations to provide more definite information on the possible cosmological nature of the hinted SGWB, extending their constraining power on the physics governing the early universe.

\acknowledgments
We thank G. Lucente, I. Musco,  D. Racco, F. Rompineve,  C. Smarra, J. Urrutia, V.Vaskonen, and H. Veerm\"ae for interesting discussions.
A.J.I. thanks the NICPB (Tallinn, Estonia) for the nice hospitality during the realisation of this project. 
The research of A.U. was supported in part by the MIUR under contract 2017\,FMJFMW (``{New Avenues in Strong Dynamics},'' PRIN\,2017).
G.F. acknowledges financial support provided under the European
Union's H2020 ERC, Starting Grant agreement no.~DarkGRA--757480 and under the MIUR PRIN programme, and support from the Amaldi Research Center funded by the MIUR program ``Dipartimento di Eccellenza" (CUP:~B81I18001170001).
This work was supported by the EU Horizon 2020 Research and Innovation Programme under the Marie Sklodowska-Curie Grant Agreement No. 101007855 and 
and  additional financial support provided by ``Progetti per Avvio alla Ricerca - Tipo 2", protocol number
AR2221816C515921.
A.J.I. acknowledges additional financial support provided under the ``Progetti per Avvio alla Ricerca Tipo 1", protocol number
AR12218167D66D36, and additional financial support provided under the ``Progetti di mobilità di studenti di dottorato di ricerca".

\appendix
\section{Fine tuning and initial conditions in the curvaton model}\label{app:Ini}
In order for the model to produce the needed enhancement of the angular perturbations, $\varphi$ has to start its rolling from some value $\varphi_* = \mathcal{O}(\bar{M}_{\rm Pl})$. Ref.\,\cite{Kasuya:2009up} justifies the Planckian initial condition by arguing the existence of some
 pre-inflationary phase during which the field $\varphi$ gets a negative Hubble-induced mass term. 
Instead of invoking some custom-made pre-inflationary physics, let us try to understand if this initial condition can be considered as a natural outcome of inflationary dynamics. 

\begin{figure}[h]
\begin{center}
$$\includegraphics[width=.49\textwidth]{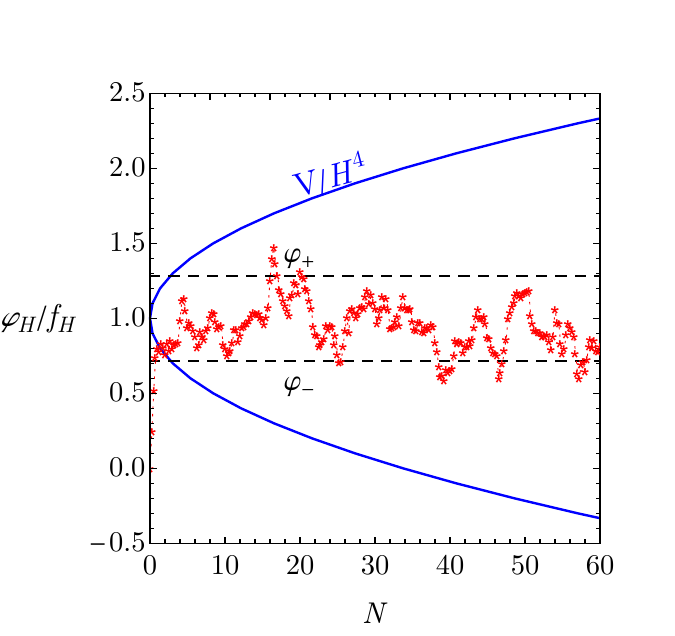}
\qquad\includegraphics[width=.49\textwidth]{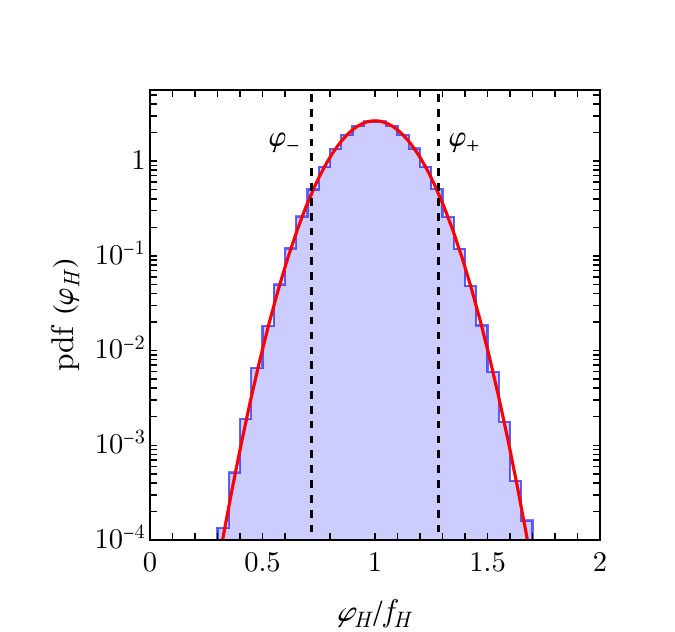}$$
\caption{\em Left panel: Red dots represent the stochastic dynamics of the field $\varphi$ in terms of the number of $e$-folds N while the blue line is the quadratic potential in equation (\ref{eq:EffectiveV}).
Right panel: The histogram represents the distribution of the different numerical solutions of equation (\ref{eq:QuantumKicks}) while the red line is the Gaussian function in equation (\ref{eq:PDFGauss}).} \label{fig:Stoca} 
\end{center}
\end{figure}

During inflation, the stochastic dynamics of the field $\varphi$ in the de Sitter background is described by the equation
\begin{align}\label{eq:QuantumKicks}
\frac{d^2\varphi_H}{dN^2} + 3\frac{d\varphi_H}{dN} + c(\varphi_H -f_H) = 
\eta_H\,,
~~~~~~~~
\langle \eta_H(N)\eta_H(N^{\prime})\rangle = \frac{9}{4\pi^2}\delta(N - N^{\prime})\,,
\end{align}
where the left-hand side of eq.\,(\ref{eq:QuantumKicks}) describes the evolution of the long-wavelength modes 
while the right-hand side represents the quantum noise sourced by the short-wavelength ones. 
Notice that this stochastic picture is applicable in the range of values $0 < c < 9/4$ that we consider in section\,\ref{sec:TypeI}.
A realization of the numerical solution of the above stochastic differential equation is shown in the left panel of fig.\,\ref{fig:Stoca}.
One can compute several times the solution of the above stochastic equation but it is very likely (in a way the we shall quantify in a moment) that the outcome will be always similar: the motion of the field $\varphi_H$ remains confined close to the minimum of the potential.  
More in detail, we can define the limiting values
\begin{align}
\varphi_{\pm} = f_H \pm \frac{3}{2\pi c}\,,
\end{align}
which correspond to the two dashed lines in fig.\,\ref{fig:Stoca}. 
The random motion of the field $\varphi_H$ does not overcome these two values. 
This is simple to understand.
When $|\varphi_H| > \varphi_{\pm}$, the classical displacement (per unit Hubble time) becomes larger than the amplitude of the quantum jump and the field $\varphi_H$ is pushed towards the minimum of the potential.
We can actually do better and compute the probability to find, after $N$ $e$-fold of inflation and at some position in space, some specific value of the field $\varphi_H$. 
This probability can be computed numerically by solving many times eq.\,(\ref{eq:QuantumKicks}) and extracting from the resulting statistical sample the corresponding  
PDF or by solving the Fokker-Planck equation. The two procedure agree, and we find that, after few $e$-folds, the PDF is well described by the Gaussian distribution
\begin{align}\label{eq:PDFGauss}
{\it pdf}(\varphi_H) = \frac{1}{\sqrt{2\pi}\sigma}\exp\left[
-\frac{(\varphi_H - f_H)^2}{2\sigma^2}
\right]\,,~~~~~~~{\rm with\,variance}~~~\sigma = \frac{1}{2\pi}\sqrt{\frac{3}{2c}}\,.
\end{align}
Clearly, the probability to find $\varphi_H$ at Planckian values is an exponentially small number and one should admit a certain degree of fine-tuning in the initial conditions of the model.

One can try to justify the small probability in eq.\,(\ref{eq:PDFGauss}) with some anthropic reasoning. 
In the context of the multiverse picture, one should multiply the small probability in eq.\,(\ref{eq:PDFGauss}) times the number of vacua in the landscape.
The latter could well be a gargantuan number, and it may turn what seems to be an extremely unlucky event (from the point of view of one single universe) 
into a plausible property of the landscape.

This perspective is acceptable as long as we link the properties of the model to some 
anthropic observable like  the ratio of dark matter to baryon matter\,\cite{Hellerman:2005yi}.

\section{Perturbations in a Friedmann-Robertson-Walker universe}\label{app:Pert}

In this appendix we review the formalism used in our analysis for the description of scalar
 perturbations in a Friedmann-Robertson-Walker universe with multiple interacting fluids. 
 We refer the reader to ref.\,\cite{Kodama:1984ziu} for a more comprehensive discussion.

The perturbed line element is
 \begin{align}\label{eq:MetricPerturbation}
ds^2 = 
&
-[1+2A(t,\vec{x})]dt^2 + 2a(t)\partial_{i}B(t,\vec{x})dx^i dt
\nonumber \\
& + 
  a(t)^2\big\{
 [1-2\psi(t,\vec{x})]\delta_{ij} + 2\partial_{ij}E(t,\vec{x})
  \big\}dx^i dx^j\,.
 \end{align} 
 Derivatives with respect to the cosmic time $t$ are indicated with a dot, $\dot{}\equiv d/dt$.
 The intrinsic curvature of a spatial hypersurface, $R_3$, is given by $R_3 = (4/a^2)\nabla^2\psi$.\footnote{
We write $ds^2 = g_{\mu\nu}dx^{\mu}dx^{\nu} = [g^{(0)}_{\mu\nu}(t) + 
\delta g_{\mu\nu}(t,\vec{x})]
dx^{\mu}dx^{\nu}$ with $g^{(0)}_{\mu\nu}(t) = {\rm diag}(-1,a(t)^2,a(t)^2,a(t)^2)$. 
Notice that in eq.\,(\ref{eq:MetricPerturbation}) the matrix $\partial_{ij}E(t,\vec{x})$ is {\it not} traceless. 
Alternatively, one can define the perturbed metric as
 \begin{align}\label{eq:MetricPerturbation2}
ds^2 = -[1+2A(t,\vec{x})]dt^2 + 2a(t)\partial_{i}B(t,\vec{x})dx^i dt + 
  a(t)^2\big\{
 [1-2D(t,\vec{x})]\delta_{ij} + 2\bar{\partial}_{ij}E(t,\vec{x})
  \big\}dx^i dx^j\,,
 \end{align} 
with $\bar{\partial}_{ij} = \partial_{ij} - (1/3)\delta_{ij}\nabla^2 E(t,\vec{x})$. 
The matrix $\bar{\partial}_{ij}E(t,\vec{x})$ is now, by construction, traceless. 
Of course, $D(t,\vec{x})$ in eq.\,(\ref{eq:MetricPerturbation2}) differs from $\psi(t,\vec{x})$ in eq.\,(\ref{eq:MetricPerturbation}). 
In particular, using eq.\,(\ref{eq:MetricPerturbation2}) the intrinsic
curvature of a spatial hypersurface is now given by $\bar{R}_3 = (4/a^2)\nabla^2(D + \nabla^2E/3)$. 
Consequently, the spatially-flat gauge is defined by $\psi = 0$ if one takes eq.\,(\ref{eq:MetricPerturbation}) but 
by $D  = -\nabla^2E/3$ if one takes eq.\,(\ref{eq:MetricPerturbation2}).
} 
We split the Einstein field equations $G_{\mu\nu}\equiv R_{\mu\nu} - \frac{1}{2}g_{\mu\nu}R = 8\pi G_N T_{\mu\nu}$ 
into a zero-order  system of equations 
$G_{\mu\nu}^{(0)} = 8\pi G_N T_{\mu\nu}^{(0)}$
describing the background dynamics (discussed in section\,\ref{sec:AfterInfla}) 
plus linear perturbations $\delta G_{\mu\nu} = 8\pi G_N \delta T_{\mu\nu}$ 
(relevant for the computations carried out in section\,\ref{eq:Per}). 
The reduced Planck mass $\bar{M}_{\rm Pl}$ is related to the Newton gravitational constant $G_N$ by 
$\bar{M}_{\rm Pl}^2 = 1/8\pi G_N$.

The stress-energy tensor of a fluid with energy density $\rho$, isotropic pressure $P$ and four-velocity $u^{\mu}$ 
is given by $T^{\mu}_{\,\,\nu} = (\rho + P)u^{\mu}u_{\nu} + P\delta^{\mu}_{\,\,\nu}$.\footnote{
We set to zero the anisotropic stress tensor. 
Scalar fields and perfect fluids cannot support anisotropic stress.} 
The total four velocity is subject to the constraint $u^{\mu}u_{\mu} = -1$. 
At the linear order in the perturbations, we have $u_{\mu} = (-(1+A), a(\partial_i v + \partial_i B))$ where 
$v$ is the total scalar velocity potential.
Consequently, we find
$T^{\mu}_{\,\,\nu} = 
{\rm diag}(-\rho,P,P,P) + \delta T^{\mu}_{\,\,\nu}$
with 
\begin{align}\label{eq:EnergyMomentumTensor}
\delta T^{0}_{\,\,0} = -\delta\rho\,,
~~~~~
\delta T^{0}_{\,\,i} = (\rho + P)a(\partial_i B + \partial_i v)\,,
~~~~~
\delta T^{i}_{\,\,0} = - \frac{(\rho + P)}{a}\partial_i v\,,
~~~~~
\delta T^{i}_{\,\,j} = \delta^{i}_{j}\,\delta P\,. 
\end{align}
The $^0_{\,\,0}$ and $^0_{\,\,i}$ components of the perturbed Einstein field equations -- that are
 $\delta G^0_{\,\,0} = 
8\pi G_N \delta T^0_{\,\,0}$ and 
$\delta G^0_{\,\,i} = 
8\pi G_N \delta T^0_{\,\,i}$, respectively -- 
 read
\begin{align}
3H(\dot{\psi} + HA) - \frac{\nabla^2}{a^2}\big[
\psi + H(\underbrace{a^2 \dot{E} - aB}_{\equiv\,\chi})
\big] 
 + 4\pi G_N \delta\rho & = 0\,,\label{eq:Funda1}\\
\dot{\psi} + HA + 4\pi G_N(\rho + P)\underbrace{a(B + v)}_{\equiv\,V} & = 0\,,\label{eq:Funda2}
\end{align}
where we defined the scalar shear $\chi \equiv a^2 \dot{E} - aB$ and the total covariant velocity perturbation 
$V\equiv a(B+v)$. 
The off-diagonal spatial components of the perturbed Einstein field equations give 
the time-evolution of the scalar shear 
\begin{align}\label{eq:EvoChi}
\dot{\chi} + H\chi - A + \psi = 0\,.
\end{align}
Finally, the spatial trace of the perturbed Einstein field equations, combined with eq.\,(\ref{eq:EvoChi}), gives 
\begin{align}
\ddot{\psi} + 3H\dot{\psi} + H\dot{A} + (3H^2 + 2\dot{H})A - 4\pi G_N \delta P = 0\,.\label{eq:Funda3} 
\end{align}

Through the Bianchi identities, the Einstein field equations 
imply the local conservation of {\it total} energy and momentum, that is $\nabla_{\mu}T^{\mu\nu} = 0$. 
In the multiple fluid case the total energy-momentum tensor is the sum of the energy-momentum tensors of the
individual fluids, $T^{\mu\nu} = \sum_{\alpha}T^{\mu\nu}_{\alpha}$. 
For each fluid, we write the local energy-momentum transfer 4-vector as 
$\nabla_{\mu}T^{\mu\nu}_{\alpha} = Q^{\nu}_{\alpha}$ with the constraint $\sum_{\alpha}Q^{\nu}_{\alpha} = 0$.
At the zero-order in the perturbations, 
the time component $\nu = 0$ of $\nabla_{\mu}T^{\mu\nu} = 0$ gives 
$\dot{\rho} + 3H(\rho + P) = 0$; at the linear order in the perturbations, on the contrary, we find
\begin{align}\label{eq:EnCon}
\dot{\delta\rho} + 3H(\delta\rho + \delta P) - 3\dot{\psi}(\rho + P) + 
\frac{\nabla^2}{a^2}\big[(\rho + P)(\chi + V)\big] = 0\,.
\end{align}
For a single fluid identified by the label $\alpha$,  
$\nabla_{\mu}T^{\mu 0}_{\alpha} = Q^{0}_{\alpha}$ gives the zero-order result
$\dot{\rho}_{\alpha}+3H(\rho_{\alpha} + P_{\alpha}) = Q_{\alpha}$ while its perturbed version reads
\begin{align}\label{eq:IndividualEnCon}
\dot{\delta\rho}_{\alpha} + 3H(\delta\rho_{\alpha} + \delta P_{\alpha}) - 3\dot{\psi}(\rho_{\alpha} + P_{\alpha}) + 
\frac{\nabla^2}{a^2}\big[(\rho_{\alpha} + P_{\alpha})(\chi + V_{\alpha})\big] - AQ_{\alpha} - \delta Q_{\alpha}= 0\,,
\end{align}
where we defined the covariant velocity perturbation of the $\alpha$-fluid as 
$V_{\alpha} \equiv a(v_{\alpha} + B)$ where $v_{\alpha}$ 
is the  scalar velocity potential for the $\alpha$-fluid. 
The total fluid perturbations are related to the individual fluid quantities by 
\begin{align}
\delta\rho \equiv \sum_{\alpha}\delta\rho_{\alpha}\,,~~~~~~~~~~
\delta P \equiv \sum_{\alpha}\delta P_{\alpha}\,,~~~~~~~~~~
V = \sum_{\alpha}\frac{\rho_{\alpha} + P_{\alpha}}{\rho + P}\,V_{\alpha}\,,
\end{align}
through which one can get eq.\,(\ref{eq:EnCon}) by summing 
eqs.\,(\ref{eq:IndividualEnCon}). In eq.\,(\ref{eq:IndividualEnCon}) $Q_{\alpha}$ is the energy transfer to the 
$\alpha$-fluid and $\delta Q_{\alpha}$ its perturbation. 
Momentum conservation, $\nabla_{\mu}T^{\mu i}_{\alpha} = Q^{i}_{\alpha}$, gives for the $\alpha$-fluid 
the equation 
\begin{align}\label{eq:IndividualMomCon}
\dot{V}_{\alpha} + \bigg[
\frac{Q_{\alpha}}{(\rho_{\alpha} + P_{\alpha})}(1+c_{\alpha}^2) - 3Hc_{\alpha}^2 
\bigg]V_{\alpha} + A + 
\frac{1}{\rho_{\alpha} + P_{\alpha}}\big(
\delta P_{\alpha} - Q_{\alpha}V
\big) = 0\,,
\end{align}
where $c_{\alpha}^2 \equiv \dot{P}_{\alpha}/\dot{\rho}_{\alpha}$ is the adiabatic sound speed of the $\alpha$-fluid. 
We consider in the above equation the case of zero momentum transfer among the fluids. 
The total momentum conservation equation $\nabla_{\mu}T^{\mu i} = 0$ is given by
\begin{align}\label{eq:MomCon}
\dot{V} - 3Hc_s^2 V + A + \frac{1}{\rho + P}\,\delta P = 0\,,
\end{align}
where $c_s^2 \equiv \dot{P}/\dot{\rho}$ is the total adiabatic speed of sound which can be written as a weighted 
sum of the adiabatic sound speeds of the
individual fluids 
\begin{align}
c_s^2 = \sum_{\alpha}\frac{\dot{\rho}_{\alpha}}{\dot{\rho}}c_{\alpha}^2\,.
\end{align}
In summary, the relevant equations are eqs.\,(\ref{eq:Funda1},\,\ref{eq:Funda2},\,\ref{eq:EvoChi}\,,\ref{eq:Funda3}) 
with the energy and momentum conservation in eqs.\,(\ref{eq:EnCon},\,\ref{eq:IndividualEnCon}) 
and eqs.\,(\ref{eq:IndividualMomCon},\,\ref{eq:MomCon}).
 
We consider the description of the dynamics of the scalar perturbations in terms of gauge-invariant quantities. 
We define {\it i)} the total curvature perturbation on
uniform density hypersurfaces
\begin{align}
\zeta \equiv -\psi - H\frac{\delta\rho}{\dot{\rho}} = \sum_{\alpha}\frac{\dot{\rho}_{\alpha}}{\dot{\rho}}\,\zeta_{\alpha}\,,
~~~~~~~{\rm with}~~~~~~\zeta_{\alpha} \equiv -\psi - H\frac{\delta\rho_{\alpha}}{\dot{\rho}_{\alpha}}\,,
\label{eq:ZetaDefinititon}
\end{align}
{\it ii)} the total comoving curvature perturbation 
\begin{align}
\mathcal{R} \equiv \psi - HV = 
\sum_{\alpha}\frac{\rho_{\alpha} + P_{\alpha}}{\rho + P}\,\mathcal{R}_{\alpha}\,,
~~~~~~~{\rm with}~~~~~~
\mathcal{R}_{\alpha} \equiv \psi - HV_{\alpha}\,,\label{eq:DDefiR}
\end{align}
and {\it iii)} the curvature perturbation on
uniform shear hypersurfaces
\begin{align}
\Psi \equiv \psi + H\chi\,,~~~~~~~{\rm with}~~~~~~\chi \equiv a^2 \dot{E} - aB\,.
\end{align}
From now on, we move to consider the dynamics in Fourier space. 
This implies the formal substitution $\nabla^2 \to -k^2$, where $k$ is the comoving wavenumber (with $k\equiv |\vec{k}|$). 
Furthermore, each perturbed quantity should be now understood as a specific Fourier mode with comoving wavenumber $k$. 
Notice that the system formed by eq.\,(\ref{eq:Funda1}) and eq.\,(\ref{eq:Funda2}), together with the 
equations governing the background dynamics, gives the relation
\begin{align}\label{eq:ZetaR}
3\dot{H}(\zeta + \mathcal{R}) = \frac{k^2}{a^2}\,\Psi\,,~~~~
{\rm or\,equivalently\,}~~~-\frac{3}{2}\big(1+P/\rho\big)(\zeta + \mathcal{R}) = \frac{k^2}{(a H)^2}\,\Psi\,,
\end{align}
which shows that on super-Hubble scale, where $k^2/(aH)^2 \ll 1$, we have $-\zeta \simeq  \mathcal{R}$. 

The evolution equation for $\zeta_{\alpha}$ can be obtained by taking the time derivative of its definition 
in eq.\,(\ref{eq:ZetaDefinititon}) and using eq.\,(\ref{eq:IndividualEnCon}) for $\dot{\delta\rho}_{\alpha}$ 
(in conjunction
with eq.\,(\ref{eq:Funda1}) and the background equations). We find
\begin{align}\label{eq:GaugeInv1}
\dot{\zeta}_{\alpha} =  
- 
\frac{\dot{H}Q_{\alpha}}{H\dot{\rho}_{\alpha}}\big(
\zeta - \zeta_{\alpha}
\big) 
 + 
 \frac{k^2}{3a^2 H}\bigg[
 \Psi - \bigg(
 1 - \frac{Q_{\alpha}}{\dot{\rho}_{\alpha}}
 \bigg)\mathcal{R}_{\alpha}
 \bigg]
 \nonumber 
 \\
 +
 \frac{3H^2}{\dot{\rho}_{\alpha}}\underbrace{\big(
\delta P_{\alpha} - c_{\alpha}^2 \delta\rho_{\alpha}
\big)}_{\equiv\,\delta P_{{\rm intr,}\alpha}}
- \frac{H}{\dot{\rho}_{\alpha}}\underbrace{\bigg(
\delta Q_{\alpha} - \frac{\dot{Q}_{\alpha}\delta\rho_{\alpha}}{\dot{\rho}_{\alpha}}
\bigg)}_{\equiv\,\delta Q_{{\rm intr,}\alpha}}
 \,,
\end{align}
where the combination $\delta P_{\alpha} - c_{\alpha}^2 \delta\rho_{\alpha}$ defines 
the so-called intrinsic non-adiabatic pressure perturbation of the $\alpha$-fluid $P_{{\rm intr,}\alpha}$.
For a barotropic fluid, that is a fluid with equation of state $P_{\alpha} = P_{\alpha}(\rho_{\alpha})$, 
the intrinsic non-adiabatic pressure perturbation vanishes since in this case 
we simply have $\delta P_{\alpha} = (\dot{P}_{\alpha}/\dot{\rho}_{\alpha})\delta \rho_{\alpha}$.

The combination 
$\delta Q_{\alpha} - \dot{Q}_{\alpha}\delta\rho_{\alpha}/\dot{\rho}_{\alpha}$ defines the so-called
intrinsic non-adiabatic energy transfer perturbations of the $\alpha$-fluid $\delta Q_{{\rm intr,}\alpha}$. 
Notice that $\delta Q_{{\rm intr,}\alpha}$ vanishes if 
the energy transfer $Q_{\alpha}$ is a function of the density $\rho_{\alpha}$ so that 
  $\delta Q_{\alpha} = (\dot{Q}_{\alpha}/\dot{\rho}_{\alpha})\delta \rho_{\alpha}$.

From the definition in eq.\,(\ref{eq:DDefiR}) 
we write $\dot{\mathcal{R}}_{\alpha} = \dot{\psi} - \dot{H}V_{\alpha} - H\dot{V}_{\alpha}$. 
The evolution equation for  $\mathcal{R}_{\alpha}$, therefore, can be obtained from the momentum conservation in 
eq.\,(\ref{eq:IndividualMomCon}). Using the background equations and eq.\,(\ref{eq:Funda1}), we find
\begin{align}\label{eq:GaugeInv2}
\dot{\mathcal{R}}_{\alpha} = 
(\mathcal{R} - \mathcal{R}_{\alpha})\bigg(
\frac{Q_{\alpha}}{\rho_{\alpha} + P_{\alpha}} - \frac{\dot{H}}{H}
\bigg)  - \frac{c_{\alpha}^2 \dot{\rho}_{\alpha}}{\rho_{\alpha} + P_{\alpha}}
\big(\zeta_{\alpha} + \mathcal{R}_{\alpha}\big) + \frac{H}{\rho_{\alpha} + P_{\alpha}}
\big(
\delta P_{\alpha} - c_{\alpha}^2 \delta\rho_{\alpha}
\big)\,.
\end{align}
Similarly, eq.\,(\ref{eq:EnCon}) gives the evolution equation for the total curvature perturbation $\zeta$, and we find
\begin{align}\label{eq:GaugeInv3}
\dot{\zeta} = -\frac{H}{\rho + P}\underbrace{\big(
\delta P - c_s^2 \delta \rho
\big)}_{=\,\delta P_{\rm nad}} + \frac{k^2}{3a^2 H}\big(\Psi - \mathcal{R}\big)\,,
\end{align}
where on the right-hand side we used the definition of the non-adiabatic pressure perturbation 
$\delta P \equiv \delta P_{\rm nad} + c_s^2 \delta \rho$. 
In the presence of more than one fluid, the total non-adiabatic pressure perturbation $\delta P_{\rm nad}$ 
consists of two parts, $\delta P_{\rm nad} \equiv \delta P_{\rm intr} + \delta P_{\rm rel}$. 
The first part is due to the intrinsic entropy perturbation of each fluid, 
$\delta P_{\rm intr} = \sum_{\alpha}\delta P_{{\rm intr,}\alpha}$ with $\delta P_{{\rm intr,}\alpha}$
as defined in eq.\,(\ref{eq:GaugeInv1}); 
the second part of the non-adiabatic pressure perturbation, $\delta P_{\rm rel}$, 
is due to the relative entropy perturbation $\mathcal{S}_{\alpha\beta}\equiv 3(\zeta_{\alpha} - \zeta_{\beta})$
between different fluids
\begin{align}
\delta P_{\rm rel} = -\frac{1}{6H\dot{\rho}}\sum_{\alpha,\beta}\dot{\rho}_{\alpha}\dot{\rho}_{\beta}
(c_{\alpha}^2 - c_{\beta}^2)\mathcal{S}_{\alpha\beta} = 
 -\frac{1}{2H\dot{\rho}}\sum_{\alpha,\beta}\dot{\rho}_{\alpha}\dot{\rho}_{\beta}
(c_{\alpha}^2 - c_{\beta}^2)(\zeta_{\alpha} - \zeta_{\beta})\,.
\end{align}

From eq.\,(\ref{eq:MomCon}), on the other hand, we get
\begin{align}\label{eq:GaugeInv4}
\dot{\mathcal{R}} = 
\bigg(
\frac{\dot{H}}{H} + 3Hc_s^2 
\bigg)\big(\zeta + \mathcal{R}\big) + \frac{H}{\rho + P}\big(
\delta P - c_{\alpha}^2 \delta\rho
\big) - \frac{k^2}{3a^2 H}\,\Psi\,.
\end{align}
\subsection{Perturbations dynamics in the Axion-curvaton model}

We now interpret eqs.\,(\ref{eq:GaugeInv1},\,\ref{eq:GaugeInv2},\,\ref{eq:GaugeInv3},\,\ref{eq:GaugeInv4}) 
in light of the curvaton model studied in the main body of this paper.  
We have two fluid species, namely the curvaton and the radiation field, that are identified, respectively, with the labels $\alpha = \phi,\gamma$. The curvaton field decays into radiation with a decay rate $\Gamma_{\phi}$, which we take to be a constant.  

Consider the dynamics during phase I. We set $\Gamma_{\phi} = 0$ so that we do not have energy transfer between the scalar field and radiation. Furthermore we neglect the curvature perturbation and comoving curvature perturbation of the radiation. Hence the eqs.\,\ref{eq:ZetaDefinititon} and eqs.\,\ref{eq:DDefiR} simply read as
\begin{align}
&\zeta_{\phi} \equiv -\psi - H\frac{\delta\rho_{\phi}}{\dot{\rho}_{\phi}} = \frac{2 \ \delta\theta_0}{3 \ \theta_0}\,,
~~~~~~{\rm and}~~~~~~\zeta \equiv  \sum_{\alpha}\frac{\dot{\rho}_{\alpha}}{\dot{\rho}}\,\zeta_{\alpha}= \frac{\dot{\rho}_{\phi}}{\dot{\rho}_{\phi}+\dot{\rho}_{\gamma}}\zeta_{\phi}\,,\label{eq:Zetaaxion}
\\
&\mathcal{R}_{\phi} \equiv \psi - HV_{\phi}=  \frac{ \delta\theta_0}{ \theta_0}\frac{H \ \theta}{\delta\dot{\theta}}\,,
~~~~~~{\rm and}~~~~~~
\mathcal{R} \equiv \sum_{\alpha}\frac{\rho_{\alpha} + P_{\alpha}}{\rho + P}\,\mathcal{R}_{\alpha}=\frac{\rho_{\phi} + P_{\phi}}{\rho_{\phi} +(4/3)\rho_{\gamma} +P_{\phi}}\mathcal{R}_{\phi}\, ,\label{eq:Raxion}
\end{align}
where in this case the time-dependent quantities are evaluated solving the system given by eqs.\,(\ref{eq:DynBGSim1}-\ref{eq:DynBGSim3}).
We can use these equations as the initial conditions for the system that describe the evolution of the perturbations during phase II and III.
For the sake of clarity, we also introduced explicitly the subscript $_k$ to remark that perturbations are Fourier modes with fixed comoving wavenumber $k$.

Now we consider the case of phase II+III, as defined in section\,\ref{sec:AfterInfla}.
The scalar field $\phi$ verifies the Klein-Gordon equation of motion $\ddot{\phi} + 3H\dot{\phi} + \mathcal{V}^{\prime}(\phi) = 0$ 
and its energy density and pressure are given by
\begin{align}\label{eq:ScalarFluid}
\rho_{\phi} = \frac{1}{2}\dot{\phi}^2 + \mathcal{V}(\phi)\,,~~~~~P_{\phi} = \frac{1}{2}\dot{\phi}^2 - \mathcal{V}(\phi)\,,
\end{align}
with adiabatic speed of sound
\begin{align}
c_{\phi}^2 = \frac{\dot{P}_{\phi}}{\dot{\rho}_{\phi}} = 1 + \frac{2\mathcal{V}^{\prime}(\phi)}{3H\dot{\phi}}\,. 
\end{align}
The energy transfer from the curvaton field to radiation is described by 
$Q_{\phi} = -\Gamma_{\phi}\rho_{\phi}$ (and, consequently, 
$Q_{\gamma} = \Gamma_{\phi}\rho_{\phi}$).
Radiation is a perfect fluid with $P_{\gamma} = \rho_{\gamma}/3$.
The perturbations in the energy transfer are described by 
$\delta Q_{\phi} = -\Gamma_{\phi}\delta\rho_{\phi}$ and 
$\delta Q_{\gamma} = \Gamma_{\phi}\delta\rho_{\phi}$ (where, as stated before, we are assuming that $\delta\Gamma_{\phi} = 0$). 
Consequently, as discussed below eq.\,(\ref{eq:GaugeInv1}), we have 
\begin{align}
\delta Q_{{\rm intr,}\phi} = 
\delta Q_{\phi} - \frac{\dot{Q}_{\phi}\delta\rho_{\phi}}{\dot{\rho}_{\phi}} = 0\,.
\end{align}
During this phase, we have $P_{\phi} = 0$, and the scalar field behaves as a pressure-less fluid. 
Consequently, we have $\delta P_{{\rm intr,}\phi} = 0$. 
Eq.\,(\ref{eq:GaugeInv1}), therefore, simplifies to 
\begin{align}
\dot{\zeta}_{\phi} =  - 
\frac{\dot{H}Q_{\phi}}{H\dot{\rho}_{\phi}}\big(
\zeta - \zeta_{\phi}
\big)
 + 
 \frac{k^2}{3a^2 H}\bigg[
 \Psi - \bigg(
 1 - \frac{Q_{\phi}}{\dot{\rho}_{\phi}}
 \bigg)\mathcal{R}_{\phi}
 \bigg]\,.
\end{align}
Using the background dynamics, and introducing the $e$-fold time as time variable, 
we recast the previous equation in the form
\begin{tcolorbox}[colframe=gray!20,colback=lightgray!10,width=1.032\textwidth]
\vspace{-.4cm}
\begin{align}\label{eq:Final1}
\left.\frac{d\zeta_{\phi,k}}{dN}\right|_{\rm phase\,II+III} = 
\frac{(3+\Omega_{\gamma})\Gamma_{\phi}}{2(3H + \Gamma_{\phi})}\,(\zeta_k - \zeta_{\phi,k}) 
+ \frac{k^2}{3(aH)^2}\,\Psi_k - \frac{k^2}{(aH)^2}\frac{H}{(3H + \Gamma_{\phi})}\,\mathcal{R}_{\phi,k},
\end{align}
\end{tcolorbox}
\noindent 
where the notation $\left.\right|_{\rm phase\,II+III}$ remarks the fact that the corresponding evolution equation 
is strictly valid during the oscillating and decaying phase. 
To close the system, 
we need the evolution of $\zeta$, $\mathcal{R}$ and $\mathcal{R}_{\phi}$ (given that $\Psi$ is related 
to $\zeta$ and $\mathcal{R}$ via eq.\,(\ref{eq:ZetaR})).
Since radiation has a well-defined equation of state, we also have $\delta P_{{\rm intr,}\gamma} = 0$. 
Consequently, we find
\begin{align}\label{eq:Pnad}
\delta P_{\rm nad} \equiv \delta P_{\rm intr} + \delta P_{\rm rel} = 
\delta P_{{\rm intr,}\gamma} + \delta P_{{\rm intr,}\phi} + \delta P_{\rm rel}
= \delta P_{\rm rel} 
= \frac{\dot{\rho}_{\gamma}\dot{\rho}_{\phi}}{3H\dot{\rho}}(\zeta_{\phi} - \zeta_{\gamma}) 
= 
\frac{\dot{\rho}_{\phi}}{3H}(\zeta_{\phi} - \zeta)\,.
\end{align}
where in the last step we used eq.\,(\ref{eq:ZetaDefinititon}).  Eq.\,(\ref{eq:GaugeInv3}) gives
\begin{tcolorbox}[colframe=gray!20,colback=lightgray!10,width=1.032\textwidth]
\vspace{-.4cm}
\begin{align}\label{eq:Final2}
\left.\frac{d\zeta_k}{dN}\right|_{\rm phase\,II+III} = \frac{(3H + \Gamma_{\phi})\Omega_{\phi}}{(3+\Omega_{\gamma})H}\,
(\zeta_{\phi,k} - \zeta_k) + \frac{k^2}{3(aH)^2}\,(\Psi_k - \mathcal{R}_k).
\end{align}
\end{tcolorbox}

The evolution of $\mathcal{R}_k$ is governed by 
\begin{tcolorbox}[colframe=gray!20,colback=lightgray!10,width=1.032\textwidth]
\vspace{-.4cm}
\begin{align}\label{eq:Final3}
\left.\frac{d\mathcal{R}_k}{dN}\right|_{\rm phase\,II+III} = 
\bigg[
\frac{1}{H}\frac{dH}{dN} + \frac{
4H\Omega_{\gamma} - \Omega_{\phi}\Gamma_{\phi}
}{
H(3+\Omega_{\gamma})
}
\bigg]\mathcal{R}_k + 
\bigg(
\frac{1}{H}\frac{dH}{dN} + 1
\bigg)\zeta_k 
\nonumber 
\\
- \frac{(3H + \Gamma_{\phi})\Omega_{\phi}}{H(3+\Omega_{\gamma})}\,\zeta_{\phi,k} - 
\frac{k^2}{3(aH)^2}\,\Psi_k,
\end{align}
\end{tcolorbox}
\noindent 
while we find for $\mathcal{R}_{\phi,k}$\footnote{Notice that 
our result for the evolution of $\mathcal{R}_{\phi}$ differs from the 
corresponding equation found in ref.\,\cite{Firouzjahi:2012iz}.
} 
\begin{tcolorbox}[colframe=gray!20,colback=lightgray!10,width=1.032\textwidth]
\vspace{-.4cm}
\begin{align}\label{eq:Final4}
\left.\frac{d\mathcal{R}_{\phi,k}}{dN}\right|_{\rm phase\,II+III} = -\bigg(
\frac{\Gamma_{\phi}}{H} + \frac{1}{H}\frac{dH}{dN}
\bigg)(\mathcal{R}_k - \mathcal{R}_{\phi,k}).
\end{align}
\end{tcolorbox}
\noindent
The system formed by eqs.\,(\ref{eq:Final1},\,\ref{eq:Final2},\,\ref{eq:Final3},\,\ref{eq:Final4}) 
is subject to the relation
\begin{tcolorbox}[colframe=gray!20,colback=lightgray!10,width=1.032\textwidth]
\vspace{-.4cm}
\begin{align}
\frac{3}{H}\frac{dH}{dN}(\zeta_k + \mathcal{R}_k) = \frac{k^2}{(aH)^2}\,\Psi_k
\end{align}
\end{tcolorbox}
\noindent
We solve numerically the evolution described by eqs.\,(\ref{eq:Final1},\,\ref{eq:Final2},\,\ref{eq:Final3},\,\ref{eq:Final4})  
, that is valid during phase II and phase III ($P_{\phi} = 0$), in order to get the correct value of the curvature perturbation $\zeta_k$ for the power spectrum defined in eq.\,(\ref{eq:FinaalPS}).

\bibliographystyle{JHEP}

\bibliography{Axion-Curvaton}

\end{document}